\documentclass[journal]{IEEEtran}
\usepackage{times,amsmath,color,amssymb,graphicx,epsfig,cite,psfrag,subfigure,algorithm,algpseudocode,balance}
\usepackage{caption}
\usepackage{subfig}
\usepackage{amsfonts,pifont,enumerate,cases}
\usepackage{mathrsfs}
\usepackage[table]{xcolor}
\usepackage{verbatim}
\usepackage{bm}
\usepackage{multicol}

\newtheorem{theorem}{\underline{Theorem}}
\newtheorem{defn}{Definition}

\DeclareMathOperator*{\argmin}{argmin}

\newcommand{\vect}{\textup{vec}}

\begin{document}
\title{A Block Sparsity Based  Estimator for mmWave Massive MIMO Channels with Beam Squint}
\author{Mingjin Wang, Feifei Gao, Mark F. Flanagan, Nir Shlezinger, and Yonina C. Eldar
\thanks {M. Wang and F. Gao are with the Institute for Artificial Intelligence,
Tsinghua University (THUAI), State Key Lab of Intelligent Technologies and
Systems, Tsinghua University, Beijing National Research Center for Information
Science and Technology (BNRist), Department of Automation, Tsinghua University, Beijing, P. R. China (email: wmj17@mails.tsinghua.edu.cn, feifeigao@ieee.org).}
\thanks{M. F. Flanagan is with the School of Electrical and Electronic Engineering
University College Dublin, Belfield, Dublin 4, Ireland  (email: mark.flanagan@ieee.org).}
\thanks{N. Shlezinger and Y. C. Eldar are with the Faculty of Mathematics and Computer Science, Weizmann Institute of Science, Rehovot, Israel (e-mail: nirshlezinger1@gmail.com; yonina@weizmann.ac.il).}
\vspace{-0.5cm}
}
\maketitle
\thispagestyle{empty}
\begin{abstract}
Multiple-input multiple-output (MIMO) millimeter wave (mmWave) communication is a key technology for next generation wireless networks.
One of the consequences of utilizing a large number of antennas with an increased bandwidth is that array steering vectors vary among different subcarriers. Due to this effect, known as  \emph{beam squint}, the conventional channel model is no longer  applicable for mmWave massive MIMO systems.
In this paper, we study channel estimation under the resulting non-standard model. To that aim, we first analyze the beam squint effect from an array signal processing perspective, resulting in a model which sheds light on the \emph{angle-delay} sparsity of mmWave transmission.  We next design a compressive sensing  based channel estimation algorithm which utilizes the shift-invariant block-sparsity of this channel model. The proposed algorithm  jointly computes the \emph{off-grid} angles, the \emph{off-grid} delays, and the complex gains of the multi-path channel.
We show that  the newly proposed scheme  reflects the mmWave channel  more accurately and results in improved performance  compared to traditional approaches.
We then demonstrate how this approach can be applied to recover both the uplink as well as the downlink channel in frequency division duplex (FDD) systems, by exploiting the angle-delay reciprocity of mmWave channels.
\end{abstract}

\vspace{-0.3cm}
\section{Introduction}\label{sec:intro}
\vspace{-0.1cm}
The proliferation of wireless services such as multimedia, virtual reality, network video, vehicular networking, and the Internet of Things gives rise to  continuously increasing  demands on the transmission rate and quality  of service. These demands include  higher throughput, shorter delays, improved connectivity, denser networks, and better user experience \cite{5G}.
To meet all these requirements, it is necessary to exploit higher frequencies, and in particular, the millimeter wave (mmWave) band, to overcome the spectral congestion of standard wireless frequency bands \cite{mmWave}.
An additional method to increase the spectral efficiency and to improve  spatial resolution is to equip the base station (BS) with a large-scale antenna array \cite{MIMO1,MIMO2,MIMO3}. This technique is commonly referred to as  \emph{massive MIMO}.
Due to the short wavelength of mmWave,
utilizing large antenna arrays is also essential for successfully implementing mmWave communications. In particular,  the increased number of antennas can be used to implement directed beamforming, thus overcoming the dominant path-loss induced at mmWave with no line-of-sight \cite{jin1,jin2}.

A large body of research has been devoted to understanding the potential and challenges associated with mmWave massive MIMO communications in recent years. Rappaport \emph{et al}. proposed a model for mmWave channels based on an extensive measurement campaign  and demonstrated that the mmWave band can effectively support high-speed data transmission \cite{measure-mmwave}.
The work  \cite{capacity-mmwave}  derived capacity bounds for mmWave massive MIMO communications  based on the model of \cite{measure-mmwave}.
Additionally, \cite{mimo-mmwave} outlined the benefits, challenges, and potential solutions associated
with cellular networks utilizing  mmWave massive MIMO technology.

In order to achieve the potential benefits of mmWave massive MIMO communications, it is critical to have accurate channel state information (CSI). However, the channel characteristics in mmWave bands are quite different from  their conventional sub-6 GHz counterparts. In particular:
1) experimental studies \cite{signal-mmwave,mm1} have shown that electromagnetic waves in mmWave bands suffer from  severe path loss and have difficulty in bypassing obstacles;
2) mmWave channels have been shown to exhibit sparsity in the angle  and delay domains, which is not encountered in  microwave frequencies \cite{mm2,mm3}.
Furthermore, due to the narrow angle spread of each cluster, the channel covariance matrices (CCMs) of the resulting channel models are typically low-rank. These properties indicate that any efficient channel estimator should be able to build upon this inherent structure.

Different low-complexity estimation algorithms have been designed to exploit the sparsity or low-rank property of the channel, including CCM based approaches \cite {CCM}, compress sensing (CS)  algorithms \cite{OnCS1,OnCS2}, and angle domain based methods \cite{hongxiang}.
In particular, under the assumption of a finite scattering environment, the work \cite{CCM} mathematically demonstrated the low-rank feature of the  CCMs in mmWave communications and proposed a joint spatial division multiplexing algorithm to reduce the effective dimensions of the channel. The work \cite{OnCS1} used the low-rank structure of the CCMs to cast the  channel estimation task into a quadratic semidefinite programming (SDP) problem, which was solved using a polynomial SDP method.
With a given unitary dictionary matrix known to the BS, \cite{OnCS2} represented a virtual channel which has a common sparsity due to the fact
that the users share the same local scatters. A joint orthogonal matching pursuit (OMP) recovery algorithm was then presented in \cite{OnCS2} to estimate the channel and reduce feedback. To exploit the angle information for sparse channel estimation, \cite{hongxiang} designed a fast discrete Fourier transform (DFT) based spatial rotation algorithm to concentrate  most of the channel power on limited DFT grids and efficiently obtain the angle information for both  frequency division duplex (FDD) as well as time division duplex (TDD) systems.
In particular, \cite{hongxiang} used an array signal processing aided channel estimation scheme, where the angle information of the user
is exploited to simplify  channel estimation.
A detailed overview on signal processing methods, including array signal processing techniques, for mmWave massive MIMO communications can be found in \cite{signal-mmwave}.

An important drawback of the sparse channel estimation approaches mentioned above stems from the fact that they use \textit{on-grid} estimation \cite{grid-mismatch}  to solve the optimization problem, namely, they divide the continuous parameter space into a finite set of grid points. The sparse channel is then estimated assuming a  discrete dictionary, resulting in increased estimation errors as the exact parameter does not necessarily lie on the discrete grid. Such grid mismatch introduces quantization error in addition to the channel recovery uncertainty, which may reduce the ability to accurately estimate the channel.

Another  drawback of previously proposed channel estimators, e.g.,  \cite{CCM,OnCS1,OnCS2,hongxiang}, is that the  massive MIMO model used is directly obtained from the conventional MIMO model. This  model only observes the phase differences but does not capture the propagation delay of the same incident signal observed at different antennas.
This effect is negligible in conventional MIMO setups with a relatively small number of antennas, however, it cannot be ignored when the antenna array grows larger
and the bandwidth increases.
This phenomenon is referred to as the \emph{spatial-wideband effect} \cite{bolei-1}. For orthogonal frequency division multiplexing (OFDM) systems, this effect makes the array response vary with frequency, causing the beams observed by the receiver to ``deviate'' as a function of frequency \cite{beamsquint1}, which is also known as the \emph{beam squint effect} (BSE) \cite{beamsquint2}.
Since mmWave communications highly rely on the precise alignment of beams between the transmitters and the receivers, the BSE may result in
severe performance degradation if not carefully treated.

The BSE was experimentally evaluated in  \cite{beamsquint3}, which measured the beam squinting range of 15 degrees with a 4-element patch array over 6GHz bandwidth with central frequency of 60GHz.
The works \cite{beamsquint4,beamsquint5} proposed to mitigate the BSE  by integrating  interconnected slow-wave structures and metamaterial cells along the array feed network.
In \cite{beamsquint6}, the authors designed a beamforming codebook to compensate for the BSE  by imposing an achievable rate  constraint.
Nevertheless, these efforts \cite{beamsquint4,beamsquint5,beamsquint6} aim only to compensate for the channel performance loss caused by the BSE, and do not provide a systematic channel estimation scheme under this effect.

In this paper, we develop a set of channel estimation algorithms for mmWave massive MIMO systems, for both the uplink channel as well as the downlink channel, accounting for the BSE.
Our proposed algorithms operate in an off-grid manner, namely, the estimated channel coefficients can take values in a continuous set.
To that aim, we first decompose the channel vectors into angle, delay, and gain parameters, and propose a model for the BSE using these parameters. Next, we demonstrate that the task of estimating angle and delay parameters can each be expressed as a block-sparse signal recovery problem using the matrix representation of the pilot sequence.
We then propose a pilot-based block-iterative gradient descent algorithm to estimate the angle and delay parameters for the uplink channel.
Based on this estimation method, we derive an efficient angle and delay pairing algorithm with low computing complexity.
By exploiting the frequency insensitivity of angle and delay, i.e.,  \emph{angle-delay reciprocity}, we extend the proposed approach and develop an effective algorithm for downlink channel estimation in FDD systems, thus tackling one of the major problems noted in the massive MIMO literature \cite{TL}.

The rest of this paper is organized as follows:
Section~II introduces the BSE in the time and angle domains, and formulates the mmWave massive MIMO-OFDM system model.
Section~III illustrates the block sparsity based uplink  angle/delay estimation algorithm  which accounts for the BSE  and shows how it can be used to reconstruct the uplink channel.
Section~IV designs a downlink channel estimation scheme for FDD massive MIMO systems with low complexity and low overhead based on the guidelines used for deriving the uplink channel estimator.
Numerical results are provided in Section V, and Section VI concludes this paper.

\textbf{Notation:} Throughout this paper, vectors and matrices are denoted by boldface lower-case and upper-case letters, respectively; transpose, conjugate, Hermitian, inverse, and
pseudo-inverse of the matrix ${\mathbf A}$ are denoted
by ${\mathbf A}^T$,${\mathbf A}^{\ast}$, ${\mathbf A}^H$,${\mathbf A}^{-1}$ and ${\mathbf A}^{\dagger}$, respectively;
$\left\|{\mathbf A}\right\|_F$ denotes the Frobenius norm of the matrix $\mathbf A$;
$[{\mathbf A}]_{i,j}$ is the $(i,j)$th entry of ${\mathbf A}$;  indices of vectors and matrices start at 0;
$[{\mathbf A}]_{i,:}$ represents the $i$th row of the matrix ${\mathbf A}$;
$[{\mathbf A}]_{:,j}$ represents the $j$th columns of the matrix ${\mathbf A}$;
$\vect({\mathbf A})$ represents column-major vectorization of the matrix ${\mathbf A}$, i.e., the operation of stacking the columns of matrix ${\mathbf A}$ to form a vector;
$\left\|\bm h\right\|_2$ denotes the Euclidean norm of the vector $\bm h$;
$\otimes$ denotes the Kronecker product, and
$\odot$ is the Hadamard product of matrices;
${\mathbf I}_N$ is the $N\times N$ identity matrix;
$\mathbb R$ and $\mathbb C$ represent the sets of real and complex numbers, respectively;
$\mathcal R(\cdot)$ is the real part of a complex number, vector or matrix;
$ \lfloor a \rfloor$ is the downward rounding operation of a  real number $a$.


\vspace{-0.2cm}
\section{Channel Model}
\vspace{-0.1cm}
\subsection{mmWave Massive MIMO Uplink Channel Model}
\vspace{-0.1cm}
We consider a mmWave massive MIMO system, focusing  on the uplink transmission.
The BS is equipped with a uniform linear array (ULA) consisting of $M$  antennas, where $M$ is a large integer, and  the antenna spacing is $d=\lambda_c^\text{ul}/2$.  Here,  $\lambda_c^\text{ul}\triangleq {c}/{f_c^{\text{ul}}}$, where $f_c^{\text{ul}}$ is the uplink center frequency, $\lambda_c^\text{ul}$ is the  wavelength, and
$c$ is the speed of light.
To  present the channel model, highlighting the spatial wideband effect, we  discuss the case of a single user with a single antenna  and assume a noiseless setup in this subsection.
The proposed model can be extended to multiple users with multiple antennas by properly adapting the arguments in the sequel.

To formulate the channel model, we use $\alpha [i]$ to denote the discrete-time baseband transmitted symbol, with symbol period $T_s$. The continuous-time baseband transmit signal $ \bar s(t)$ can thus be expressed as
\begin{equation}\label{base band st}
  \bar s(t)=\sum_{i=-\infty}^{+\infty}\alpha[i]g(t-iT_s),
\end{equation}
where $g(t)$ is the pulse shaping function.
After modulating, the passband transmit signal $\tilde  s(t)$ can be written as

\begin{equation}\label{pass band s}
  \tilde  s(t)=\mathcal R\left\{\bar s(t)e^{j2\pi f_c^\text{ul}t}\right\}.
\end{equation}
\begin{figure}[t]
\centering
\hspace{-7mm}
\vspace{-2mm}
\includegraphics[width=90mm]{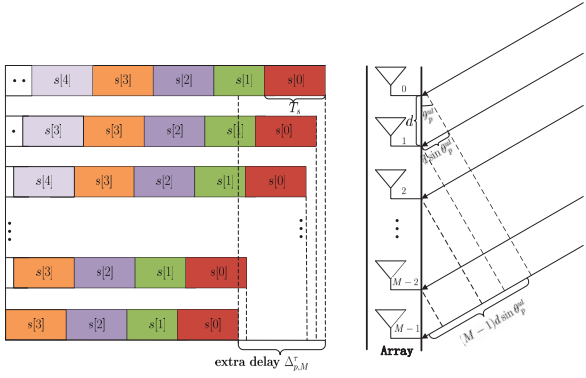}
\caption{Illustration of the non-negligible delays between the signals arriving at different antennas.
\label{fig:wideband}}
\end{figure}

\vspace{-2mm}
Let $P$ denote the number of paths between the user and the BS.  Each path has direction of arrival (DOA) $\theta_{p}^\text{ul}\in [-\pi/2,\pi/2)$ and  passband gain $\beta_{p}^\text{ul}\in \mathbb R^{+} $.  Denote the $p$th path delay between the transmitter and the first receive antenna  by $\tau_{p}^\text{ul}$. Unlike  conventional MIMO models, the large array aperture of massive MIMO receivers results in  non-negligible delays among different antennas. Those delays are present even for received signal components corresponding to  the same channel path, as illustrated in Fig. \ref{fig:wideband}. The extra delay of the $p$th path from the $m$th receive antenna  compared to the first receive antenna is given by
\begin{align}\label{tau_m}
\Delta^{\tau}_{p,m}= \frac{md\cdot\sin\theta^\text{ul}_{p}}{c}=\frac{md\cdot\sin\theta^\text{ul}_{p}}{\lambda_c^\text{ul}f_c^{\text{ul}}}.
\end{align}
Consequently, the  passband receive signal at the $m$th antenna can be written as
\begin{align}\label{pass band vander}
 &\tilde y_m^{\text{ul}}(t)=\sum_{p=1}^{P} \beta_{p}^\text{ul} \tilde s(t-\tau_{p}^\text{ul}-\Delta^{\tau}_{p,m})\notag \\
 &=\sum_{p=1}^{P}  \mathcal R\left\{\beta_{p}^\text{ul} \bar s(t-\tau_{p}^\text{ul}-\Delta^{\tau}_{p,m} )e^{j2\pi f_c^\text{ul}(t-\tau_{p}^\text{ul}-\Delta^{\tau}_{p,m})}\right\}.
\end{align}

For conventional MIMO systems, where $M$ is small or the bandwidth is narrow, it  typically holds that $\Delta^{\tau}_{p,m}\ll T_s$   for each $m \in \{0,1,\ldots,M-1\}$.  Thus the signals at different antennas satisfy $\bar s(t-\tau_{p}^\text{ul}-\Delta^{\tau}_{p,m})\approx \bar s(t-\tau_{p}^\text{ul})$. Namely, different antennas at the BS effectively observe a synchronized signal. In such scenarios, the standard MIMO channel output model, i.e., a linear convolution between the individual channel impulse responses and the same source signal, faithfully represents the received signal.

However, for  mmWave massive MIMO systems, $\Delta^{\tau}_{p,m}$ cannot be ignored for larger values of $m$, and the approximation $\bar s(t-\tau_{p}^\text{ul}-\Delta^{\tau}_{p,m})\approx \bar s(t-\tau_{p}^\text{ul})$ no longer holds. In this case, the signal observed by the first antenna will include a different time shift compared to the signals observed by other antennas. This phenomenon is referred to as the \textit{spatial  wideband effect} \cite{bolei-1}. As an illustrative example, consider a massive MIMO system  with $M=128$ BS antennas, $\theta^\text{ul}_{p}=60^\circ$, baseband symbol rate $f_s=\frac{1}{T_s}=2$GHz, and  mmWave carrier frequency $f_c^\text{ul}=60$GHz. Under this setting, the signal delay between the first antenna and the last one, computed via (3), is $1.85T_s$, which is clearly non-negligible.

In the presence of the spatial wideband  effect, it is difficult to formulate a unified discrete-time MIMO channel model since the signals arrive at different antennas with relative delays which are fractions of the symbol period.
To derive a convenient  model which facilitates analysis, we observe  the transmission in an antenna-by-antenna manner. From \eqref{pass band vander}, removing the center frequency $f_c^{\text{ul}}$, to switch from passband to baseband representation,  the continuous-time baseband receive signal at the $m$th antenna is given by
\begin{align}\label{channel ht}
 &\bar y_{m}^\text{ul}(t)  =\sum_{p=1}^{P}\beta_{p}^\text{ul} \bar s(t-\tau_{p}^\text{ul}-\Delta^{\tau}_{p,m}) e^{j2\pi f_c^\text{ul}(-\tau_{p}^\text{ul}-\Delta^{\tau}_{p,m})} \notag\\
 &= \bigg(  \sum_{p=1}^{P} \bar \beta_{p}^\text{ul} e^{-j2\pi f_c^\text{ul}\Delta^{\tau}_{p,m}}\delta(t-\tau_{p}^\text{ul}-\Delta^{\tau}_{p,m}) \bigg)  *\bar s(t) ,
\end{align}
where $\bar \beta_{p}^\text{ul} \triangleq \beta_{p}^\text{ul} e^{-j2\pi f_c^\text{ul}\tau_{p}^\text{ul}}$ is the equivalent complex channel gain, and $*$ denotes the convolution operator.
Taking the Fourier transform of \eqref{channel ht}, we obtain the  frequency-domain representation of the received signal at the $m$th antenna as:
\begin{align}\label{channel hf}
 &y_{m}^\text{ul}(f)=\int_{-\infty}^{+\infty} \bar y_{m}^\text{ul}(t)e^{-j2\pi ft}dt\notag\\
                 &=  \bigg(\sum_{p=1}^{P} \bar \beta_{p}^\text{ul} e^{-j2\pi f_c^\text{ul}\Delta^{\tau}_{p,m}} e^{-j2\pi f (\tau_{p}^\text{ul}+\Delta^{\tau}_{p,m})} \bigg) {s}(f)\notag\\
  &=\bigg(\sum_{p=1}^{P} \bar \beta_{p}^\text{ul}  e^{-j (m-1)\phi_{p}^\text{ul}}  e^{-j 2\pi f \Delta^{\tau}_{p,m}}e^{-j2\pi f \tau_{p}^\text{ul}} \bigg) {s}(f),
\end{align}
where $\phi_p^\text{ul}\triangleq \frac{2\pi d\cdot\sin(\theta_p^\text{ul})}{\lambda_c^\text{ul}} \in[-\pi, \pi)$ is defined as the normalized DOA.
For clarity, in the rest of the paper, the term ``DOA'' will refer to the normalized DOA.

By stacking the received signal $y_{m}^\text{ul}(f)$ over all $M$ antennas into a single vector representation, we can write
\begin{align}\label{channel yf_wide}
 \bm y^\text{ul}(f)= [ y_{1}^\text{ul}(f),y_{2}^\text{ul}(f), \cdots ,y_{M}^\text{ul}(f) ]^T=\bm h^\text{ul}(f)   {s}(f),
\end{align}
where
\vspace{-0.2cm}
\begin{align}\label{channel hf_wide}
 \bm h^\text{ul}(f)=\sum_{p=1}^{P} \bar \beta_{p}^\text{ul} \bm a(\phi_{p}^\text{ul},f) e^{-j2\pi f \tau_{p}^\text{ul} },
\end{align}
is the uplink channel frequency response, and
\begin{align}\label{a_theta}
\bm a(\phi_{p}^\text{ul},f)&=\left[1,\cdots,e^{-j (M-1)\phi_{p}^\text{ul}}e^{-j 2\pi f \Delta_{p,M}^{\tau}}\right]^T\notag\\
&=\left[1,\cdots,e^{-j (M-1)(1+\frac{f}{f_c^\text{ul}})\phi_{p}^\text{ul} }\right]^T,
\end{align}
is the wideband array steering vector. Note that unlike  conventional array steering vectors, see, e.g., \cite{OnCS1,OnCS2,hongxiang}, the wideband array steering vector $\bm a(\phi_p^\text{ul},f)$ is \emph{frequency-dependent}.
\begin{figure}[tb]
\centering
\vspace{-0.6cm}
\includegraphics[scale=0.32]{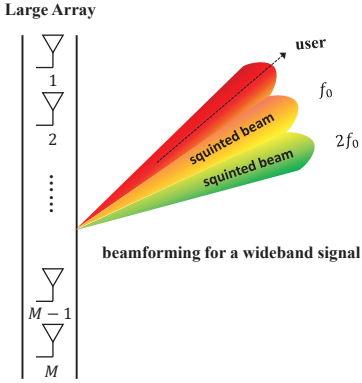}
\vspace{-0.3cm}
\caption{Illustration of the beam squint effect  directing different beam directions with the same angle.
\label{fig:beamsquint}}
\end{figure}
\vspace{-0.2cm}
\subsection{Beam Squint Effect in OFDM Systems}
\vspace{-0.1cm}
We next focus on  OFDM  signaling with $N$ subcarriers and subcarrier spacing $f_0$.
We henceforth assume that the number of antennas $M$ satisfies $M \le 2N\frac{f_c^{\rm up}}{f_s}$. This assumption is reasonable for mmWave systems in which the bandwidth is very large, thus the number of subcarriers $N$ is of the same order as $M$, and the carrier frequency $f_c^{\rm up}$ is much larger than the symbol rate $f_s$.

Using \eqref{channel hf_wide}, the uplink channel vector corresponding to the $n$th subcarrier can be written as
\vspace{-3mm}
\begin{align}\label{ofdm h-vec}
&\bm h^\text{ul}(nf_0)=\sum_{p=1}^{P} \bar \beta_{p}^\text{ul} \bm a(\phi_{p}^\text{ul},nf_0) e^{-j2\pi nf_0 \tau_{p}^\text{ul} }\notag \\
&=\sum_{p=1}^{P} {\bar \beta}_{p}^\text{ul} \left[\bm a(\phi_{p}^\text{ul},0) e^{-j2\pi nf_0 \tau_{p}^\text{ul} } \right] \odot [\bm W(\phi_{p}^\text{ul})]_{:,n},
\end{align}
for each $n=\{0,1,\ldots,N-1\}$, where $\bm W (\phi_{p}^\text{ul})$ is  the $M\times N$ matrix of wideband factors whose $(m,n)$th entry is given by
\begin{equation}\label{W_mn}
[\bm{W}(\phi_{p}^\text{ul})]_{m,n}\triangleq \text{exp}\Big(-j m\frac{nf_0}{f_c^\text{ul}}\phi_p^\text{ul}\Big).
\end{equation}
The overall $M\times N$ uplink channel matrix for the OFDM system can then be written as
\begin{align}\label{ofdm h-wide}
\mathbf H^\text{ul}=\left[ \bm h^\text{ul}(0),\bm h^\text{ul}(f_0), \cdots ,\bm h^\text{ul}((N-1)f_0) \right].
\end{align}

In previous works, e.g., \cite{OnCS1,OnCS2,hongxiang}, which do not account for the spatial-wideband effect, the factor $\Delta^{\tau}_{p,m}$ in \eqref{a_theta} is assumed to be zero. In this case, the array steering vector of any subcarrier reduces to $\bm a(\phi_{p}^\text{ul},0)$, namely, it is  \emph{frequency-independent}.
When the frequency dependence of the array steering vector is not accounted for, the postulated beam direction at the $n$th subcarrier is assumed to be  $\bm a(\phi_{p}^\text{ul},0)$, which has a phase offset of $nf_0$  compared to the true beam direction $\bm a(\phi_{p}^\text{ul},nf_0)$. This causes  a deviation between the spatially oriented beam and the user's true direction, commonly referred to as beam squint.  This deviation can lead to significant performance degradation. Fig. \ref{fig:beamsquint} depcits an example of beam squint where the beams at different subcarriers point in different directions relative to the same user angle.

\vspace{-0.2cm}
\subsection{Channel Characteristics due to the  BSE}
\vspace{-0.1cm}
We now discuss the unique channel characteristics which arise in the presence of the BSE, focusing on OFDM signaling. It follows from \eqref{ofdm h-vec} that $\bm H^\text{ul}$ can  be modeled as the sum of the contributions from $P$ paths via
\begin{align}\label{ofdm h-p}
\mathbf H^\text{ul}    &= \sum_{p=1}^{P} {\bar \beta}_{p}^\text{ul} \left[\mathbf a(\phi_{p}^\text{ul},0) \mathbf b^T(\tau_{p}^\text{ul})\right] \odot \mathbf W(\phi_{p}^\text{ul})\notag\\
                  &= \sum_{p=1}^{P} {\bar \beta}_{p}^\text{ul} \bm {\Xi}(\phi_{p}^\text{ul},\tau_{p}^\text{ul}),
\end{align}
where $\bm b(\tau_{p}^\text{ul})$ is an ${N\times 1}$ vector defined as
\begin{equation}\label{b_tau}
  \bm b(\tau_{p}^\text{ul})\triangleq
\left[1,e^{-j2\pi f_0\tau_{p}^\text{ul}},\cdots,e^{-j2\pi (N-1)f_0\tau_{p}^\text{ul}}\right]^T,
\end{equation}
and $\bm {\Xi}(\phi_{p}^\text{ul},\tau_{p}^\text{ul})\triangleq  \left[\mathbf a(\phi_{p}^\text{ul},0) \mathbf b^T(\tau_{p}^\text{ul})\right]\odot \mathbf W(\phi_{p}^\text{ul}) \in \mathbb C^{M\times N}$.

Using these notations, we state the following asymptotic channel characteristic taken from \cite{bolei-1}, which holds when the number of antennas and the number of the subcarriers both grow arbitrarily large.
\begin{theorem}
If the conditions $\frac{d\cdot f_s}{\lambda_{c}^{\text{up}}\cdot f^{\text{up}}_c}<1$ and $\frac{M-1}{2N}\frac{f_s}{f^{\text{up}}_c}<1$ are both satisfied,
then, as $M \rightarrow \infty$ and $N \rightarrow \infty$, the following property holds \cite{bolei-1}
\begin{align}\label{MN}
&\lim_{M,N\to\infty} \frac{1}{MN}\vect(\bm {\Xi}(\phi_{p}^\text{ul},\tau_{p}^\text{ul})^H\vect(\bm {\Xi}(\phi_{s}^\text{ul},\tau_{s}^\text{ul})) \notag\\
&= \bigg\{
\begin{array}{ll}
1, &\ \textup{if}\ (\phi_{p}^\text{ul},\tau_{p}^\text{ul})=(\phi_{s}^\text{ul},\tau_{s}^\text{ul}) \\
0,  &\ \text{otherwise.}
\end{array}
\end{align}
\end{theorem}
The first condition in Theorem 1 commonly holds, since  the antenna spacing $d$ is typically half of the wavelength $\lambda$, and the symbol rate $f_s$, which is typically determined by the signal bandwidth, is far less
than its carrier frequency in mmWave massive MIMO systems.
The second condition holds under our system stated in the previous subsection.

Theorem 1 implies that in the massive MIMO regime with a sufficiently large number of subcarriers, paths with different angles or delays can
be distinguished easily. We will exploit this property in Section III where we consider the reconstruction of the uplink channel.

For mmWave communications, the number of significant paths is typically much smaller compared to that encountered in standard sub 6-GHz systems \cite{mm2,mm3}, and thus  $\bm H^\text{ul}$ can be represented by a few steering vectors in the spatial and delay domains.
This sparse property indicates that CS methods can be utilized to efficiently estimate the channel parameters, as we show in the following section.

\vspace{-0.2cm}
\section{ Uplink Channel Estimation}
\vspace{-0.1cm}
In this section we propose algorithms for estimating uplink mmWave massive MIMO channels, accounting for the BSE. In particular, we  assume that the BS serves $K$ users using a MIMO-OFDM protocol \cite{pilots}, in which channel estimation is carried out using dedicated pilots in an FDD manner.
Non-overlapping subcarriers are assigned to different users.
To model the observed signal used for channel estimation, we let $\mathcal{N}_k$ denote the set of subcarrier frequencies utilized by the $k$th user, and use
$T\triangleq N/K$ to denote its cardinality\footnote{Here, we only use one OFDM block to estimate the uplink channel parameters, and  $T$  is an integer not larger than the channel coherence time. When $N$ is not an integer multiple of $K$, we use $T = \lfloor \frac{N}{K} \rfloor$. If $K$ is too large such that the block length  $T$ is not long enough to estimate the multiple channels,  one can utilize multiple OFDM blocks for channel estimation.  }. Since all subcarriers are allocated among the users without spectral overlapping, it holds that $\mathcal N_k \bigcap \mathcal N_l\! =\varnothing, k\neq l$, and that
\begin{align}\label{pilot}
\mathcal N_1 \bigcup \cdots \bigcup \mathcal N_K\!\! =\!\{0,1,\cdots, N-1\}.
\end{align}

While in the previous section we focused on a single user,  here we consider multiple users. Therefore, we henceforth use the notation $\bm H^\text{ul}_k$ to denote the channel of the $k$th user, similarly to \eqref{ofdm h-wide}.
The a-priori known pilot sequence transmitted by the $k$th user is denoted by $\bm s_k^\text{ul}\in \mathbb C^{T\times 1}$, and is assumed to have non-zero entries. This assumption accommodates a broad range of pilot sequences used in practice, such as   Zadoff-Chu (ZC) sequences \cite{3GPP}.
The received  pilots  from the $k$th user at the BS, aggregated over the corresponding subcarriers assigned to the $k$th user, can be expressed as
\vspace{-1mm}
\begin{align}\label{Y}
\bm Y_{\mathcal N_k} &=  \bm H^\text{ul}_{\mathcal N_k} \bm S^\text{ul}_{k} +\bm E_{\mathcal N_k},
\end{align}
where $\bm H^\text{ul}_{\mathcal N_k}  \in \mathbb{C}^{M\times T}$ is the subset of columns of $\bm H^\text{ul}_k$ with column indices in $\mathcal N_k$,
$\bm S^\text{ul}_{k} \triangleq \text{diag} \left\{ \bm s_k^\text{ul} \right\} \in \mathbb C^{T\times T}$ is the diagonal $k$th user pilot matrix,
and $\bm E_{\mathcal N_k}$ is  additive noise with i.i.d. zero-mean unit variance proper complex Gaussian entries.

Our goal is to reconstruct the complete channel of the $k$th user,  $\bm H_k^\text{ul}$,
from the channel output $\bm Y_{\mathcal N_k}$.  To that aim, we first identify the sparse characteristics of the unknown channel in Subsection III-A. Then, in Subsections III-B and III-C we exploit this sparse nature to efficiently estimate the unknown channel DOAs and delays, respectively. Finally, in Subsection III-D we show how these estimates can be combined to recover the unknown channel.

Since the  pilot symbols of different users do not overlap in frequency, it follows from (\ref{Y}) that the channel estimation procedure can be carried out individually for each user. Therefore, for clarity, in the rest of Section III and in Section IV, we omit the user index $k$.

\vspace{-0.2cm}
\subsection{Sparse Representation}
\vspace{-0.1cm}
To model the sparse nature of the mmWave channel coefficients with the BSE, we let the set of utilized subcarriers be written as $\mathcal{N} = \{ n_1, n_2, \ldots, n_T \}$. It follows from \eqref{ofdm h-vec} that the $q$th  column of $\bm H_{\mathcal N}^\text{ul}$, $q\in \{1,2, \cdots, T\}$, can be written as
\begin{align}
[\bm H_{\mathcal N}^\text{ul}]_{:,q}
&=\sum_{p=1}^{P} \bar \beta_{p}^\text{ul} \bm a(\phi_{p}^\text{ul},n_qf_0) e^{-j2\pi n_q f_0 \tau_{p}^\text{ul} }\notag\\
&=\sum_{p=1}^{{P}} \bar z_{q,p}^\text{ul} \bm a( \phi_{p}^\text{ul},n_qf_0),
\end{align}
where  $\bar z_{q,p}^\text{ul} =\bar \beta_{p}^\text{ul}e^{-j2\pi  n_q f_0 \tau_{p}^\text{ul} }$.

We assume that the number of possible paths, denoted by $L$, is a relatively large number, and is much larger than the number of actual paths ($L\gg {P}$).
We  define
\begin{align}\label{mmv_A}
\bm A(\bm \vartheta^\text{ul}, n_qf_0)
\triangleq[\bm a(\vartheta_1^\text{ul},n_qf_0),\cdots,\bm a(\vartheta_L^\text{ul},n_qf_0)],
\end{align}
as an \emph{overcomplete sub-dictionary} based on \eqref{a_theta}, where $\bm \vartheta^\text{ul}=[\vartheta_1^{\text{ul}},\vartheta_2^{\text{ul}},\cdots,\vartheta_L^{\text{ul}}]$ and $\vartheta_i^{\text{ul}}\triangleq {-\pi}+\frac{2i\pi}{L},  i\in\{1,2,\cdots,L\}$ divides the continuous angle space  uniformly.
Since $L$ is large, the true DOA angles
$\bm { \phi}^\text{ul}\triangleq[\phi_1^\text{ul},\cdots,\phi_P^\text{ul}]$  can be approximated (with some quantization error) to be a subset of $\bm \vartheta^\text{ul}$. This indicates that we can use the overcomplete sub-dictionary $\bm A(\bm \vartheta^\text{ul}, n_qf_0)$ to represent $[\bm H_{\mathcal N}^\text{ul}]_{:,q}$ as
\begin{align}\label{mmv_11}
[\bm H_{\mathcal N}^\text{ul}]_{:,q}\approx\bm A(\bm \vartheta^\text{ul}, n_qf_0) \bm z_q,
\end{align}
where $\bm z_q $ is a ${L\times 1}$ sparse vector whose $i$th element is
\begin{align}\label{z_s}
\hspace{-0.1cm}
z_{q,i}\!= \!\bigg\{
\begin{array}{ll}
\bar z_{q,p}^\text{ul}, &i\!=\! \argmin\limits_{ k\in \{1,\ldots,L\}} |\vartheta^\text{ul}_{k} \!-\! \phi^\text{ul}_p| ,\hspace{0.2cm} p={1,\ldots,{P}}\\
0,  &\ \text{otherwise.}
\end{array}
\end{align}
Specifically,  for each $q \in \{1,2,\ldots, T\}$,  $\bm z_q$ has at most ${P}$ nonzero values, i.e., it is   \emph{${P}$-sparse}.
It follows from \eqref{mmv_11} that $\mathbf H_{\mathcal N}^\text{ul}$ can be approximated as
\begin{align}\label{mmv_repre}
\bm H^\text{ul}_{\mathcal N}  &\approx \left[\bm A(\bm \vartheta^\text{ul},n_1f_0) \bm z_1, \cdots ,\bm A(\bm \vartheta^\text{ul}, n_Tf_0)\bm z_{T} \right].
\end{align}

We note that when the BSE is not present, the term $\bm a(\vartheta_i^\text{ul},n_qf_0)$ reduces to $\bm a(\vartheta_i^\text{ul},0)$, which is \emph{frequency-independent}. In this case,
$\bm H^\text{ul}_{\mathcal N}$ can be written as:
\begin{align}\label{mmv_narrow}
 \bm H^\text{ul}_{\mathcal N}  &\approx \left[\bm A(\bm \vartheta^\text{ul},0) \bm z_{1},\bm A(\bm \vartheta^\text{ul},0) \bm z_{2}, \cdots ,\bm A(\bm \vartheta^\text{ul},0)\bm z_{T} \right]\notag\\
   &= \bm A(\bm \vartheta^\text{ul},0)[\bm z_1,\bm z_2,\cdots,\bm z_{T}] =  \bm A(\bm \vartheta^\text{ul},0)\bm Z,
\end {align}
where $\bm Z=[\bm z_1,\bm z_2,\cdots,\bm z_T]$.
From \eqref{z_s} it holds that the sparsity pattern of ${\bm z}_q$ is independent of $q$, i.e., all the vectors $\{\bm z_q\}_{q=1}^T$ have their non-zero elements in the same  entries, so that the matrix $\bm Z$ has at most ${P}$ nonzero rows occuring on a common  index set.
Consequently, rather than trying to estimate the channel parameters from each subcarrier independently, the parameters can be jointly estimated by combining all the subcarriers,  namely, by recasting the estimation of the mmWave channel as a \textit{multiple measurement vector} (MMV) \textit{problem} \cite{eldar}.
To exploit the common support structure of $\bm Z$, one can stack the rows of $\bm Z$ and $\bm H_{\mathcal N}^\text{ul}$ into vectors. Then, \eqref{mmv_narrow} can be converted into a sparse recovery problem:
\begin{align}\label{block_1}
\text{vec}((\bm {H}_{\mathcal N}^\text{ul})^T)\approx(\bm A(\bm\vartheta^\text{ul},0)\otimes \bm I_T) \text{vec} (\bm Z^T).
\end{align}

\begin{figure}[tb]
      \centering
      \vspace{-0.6cm}
     \includegraphics[scale=0.65]{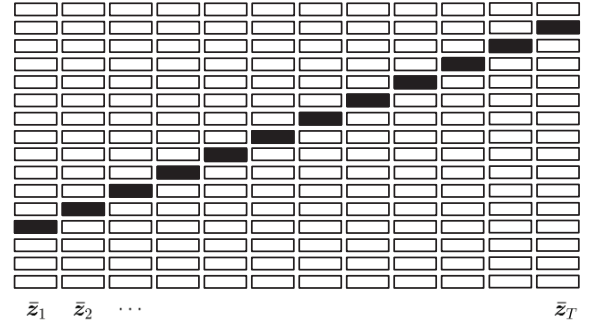}
     \caption{ Example of the structure of $\bm {\bar{Z}}$ for $P$=1.}
   \label{fig:mmv}
\end{figure}

In the presence of the  BSE, $\bm A(\bm \vartheta^\text{ul},n_qf_0)$ varies from subcarrier to subcarrier. In this case, we cannot directly obtain an expression of the form \eqref{mmv_narrow}.
Nevertheless, given that $\bm A(\bm\vartheta^\text{ul} , n_qf_0)$ has a fixed phase offset ${n_qf_0}$ compared to $\bm A(\bm\vartheta^\text{ul} , 0)$, we can write
$\bm H _{\mathcal N}^{\text{ul}}\approx\bm A(\bm \vartheta^\text{ul} ,0)\bm {\bar Z}$, where the nonzero columns of $\bm {\bar{Z}}$ have a regular `shift' characteristic.
This shift property is illustrated in Fig. \ref{fig:mmv}, in which each square corresponds to a vector entry: black squares represent the nonzero elements while blank squares indicate zeros.
Therefore, to ensure that each column of $\bm Z$ still has the same nonzero positions, we propose to  design a shift-invariant transform such that the common sparse support of the transformed $\bm Z$ satisfies   the same sparsity pattern behavior as in the absence of beam squint.

To that aim, we first define the $MT \times LT$ matrix
\begin{align}\label{D_bs}
\bm D_\text{bs}(\bm{\vartheta}^\text{ul})&\triangleq \big[ \bm D({\vartheta_1^\text{ul}}),\bm D({\vartheta_2^\text{ul}}),\cdots,\bm D({\vartheta_L^\text{ul}}) \big],
\end{align}
where  the subscript ``\emph{bs}'' stands for beam squint. Here, $\bm D({\vartheta_i^\text{ul}})$ is an $MT\times T$ matrix given by
\begin{align}\label{Dbi}
    \bm D({\vartheta_i^\text{ul}})
    &\!=\begin{bmatrix}
     [\bm a({\vartheta_i^\text{ul}},0)]_1\cdot  \bm I_T \\
     [\bm a({\vartheta_i^\text{ul}},0)]_2\cdot  \bm I_T  \\
     \vdots           \\
     [\bm a({\vartheta_i^\text{ul}},0)]_M\cdot  \bm I_T \\
    \end{bmatrix}\odot
    \begin{bmatrix}
     \bm \Phi_1(\vartheta_{i}^\text{ul}) \\
     \bm \Phi_2(\vartheta_{i}^\text{ul})   \\
     \vdots           \\
     \bm \Phi_M(\vartheta_{i}^\text{ul}) \\
    \end{bmatrix}
    \notag\\
    &\!=[\bm a(\!{\vartheta_i^\text{ul}},0)\!\otimes \!\bm I_T ]\odot \!\!\big[\bm \Phi_1^T(\vartheta_{i}^\text{ul}),\ldots,\!\bm \Phi_M^T(\vartheta_{i}^\text{ul})\big]^T\!,
\end{align}
and $\bm \Phi_m(\vartheta_{i}^\text{ul})$ is a $T\times T$ frequency rotation matrix with
parameter $\vartheta_{i}^\text{ul}$ which can be expressed as
\begin{align}\label{PHI}
&\bm \Phi_m(\vartheta_{i}^\text{ul})\triangleq
\text{diag}\Big\{e^{-j(m-1) \frac{n_1f_0}{f_c^\text{ul}} \frac{2\pi d\sin(\vartheta_i^\text{ul})}{\lambda_c^\text{ul}}},
e^{-j(m-1)}\times\notag\\
& \qquad e^{\frac{n_2f_0}{f_c^\text{ul}} \frac{2\pi d\sin(\vartheta_i^\text{ul})}{\lambda_c^\text{ul}}},\cdots ,e^{-j(m-1)   \frac{n_Tf_0}{f_c^\text{ul}}\frac{2\pi d\sin(\vartheta_i^\text{ul})}{\lambda_c^\text{ul}}  }\Big\}.
\end{align}

The application of the rotation matrix allows to express $\text{vec}((\bm H_{\mathcal N}^\text{ul})^T)$ using the matrix $\bm Z$, similarly to \eqref{block_1}.
Specifically, it can be verified that $\bm H_{\mathcal N}^\text{ul}$ is transformed into  an MMV sparse representation given by
\begin{align}\label{block_2}
\text{vec}((\bm { H}_{\mathcal N}^\text{ul})^T)
&\approx\bm D_\text{bs}(\bm \vartheta^\text{ul}) \text{vec} (\bm Z^T),
\end{align}
where, as in (22)-(23), the approximation in (28) stems from the fact that the DOAs do not necessarily lie on the grid ${\bm \vartheta^\text{ul}}$.

We henceforth refer to  $\bm D_\text{bs}(\bm \vartheta^\text{ul})$ as the \textit{sensing matrix} \cite{sm}, as it  represents a linear dimension reduction of $\text {vec} ({\bm Z}^T )$.
Since $\bm Z$ is ${P}$-row sparse, $\bm Z^T$ will be ${P}$-column sparse. This in turn implies that $\text {vec} ({\bm Z}^T )$ is
\textit{${P}$-block sparse} \cite{block}, which facilitates its recovery using block sparsity methods, as detailed in the following subsection.

\vspace{-0.2cm}
\subsection{Off-grid DOA Estimation Algorithm}
\vspace{-0.1cm}
We next study the recovery of the DOA vector from the channel output at the observed subcarriers  $\bm Y_{\mathcal N}$,  given in \eqref{Y}. Define ${\bm x} \triangleq  \text{vec} \big( (\bm Z\bm S^{\text{ul}})^T \big)$, recalling that
$ \bm S^{\text{ul}}$ is the  pilot matrix with its diagonal elements being the priori known pilot sequence.
Since the pilot elements are non-zero, $\bm x $ exhibits the same block-sparse structure as $\text {vec} ({\bm Z}^T )$.
Also, since the pilot matrix $ \bm S^{\text{ul}}$ is diagonal, we can  formulate the channel output  $\bm Y_{\mathcal N}$ as an MMV sparse representation via
\begin{align}\label{block_3}
\text{vec}((\bm {Y}_{\mathcal N})^T)
&\approx\bm D_\text{bs}(\bm \vartheta^\text{ul}) \text{vec} \big( (\bm Z\bm S^{\text{ul}})^T \big)+ \vect(\bm E_{\mathcal N}^{T})\notag\\
&=\bm D_\text{bs}(\bm \vartheta^\text{ul})\bm x +\vect(\bm E_{\mathcal N}^{T}).
\end{align}

Previously proposed algorithms for channel estimation in massive MIMO systems \cite{OnCS1,OnCS2} assume that the actual DOA values, represented by the entries of the vector $\bm\phi^\text{ul}$, coincide with values in the grid vector $\bm \vartheta^\text{ul}$. Namely, the DOA values lie on the discrete grid, and there is a one-to-one correspondence between the non-zero indexes of $\bm x$ and $\bm \phi^\text{ul}$. Then, the DOA vector is recovered from the estimated $\bm x$.
We henceforth refer to such methods as \emph{on-grid} algorithms.
However, since $\bm \phi^\text{ul}$ typically takes values in some continuous non-countable set,  the resolution of on-grid DOA estimation is only $(\frac{2\pi}{L})$,
which is also known as \emph{grid mismatch}. This grid mismatch induces quantization error and degrades  the estimation accuracy.
Although the resolution of DOA estimation can be improved by increasing $L$, denser grids implies higher, possibly non-feasible, and computational complexity.

To circumvent the grid mismatch, \emph{off-grid} solutions have been broadly studied \cite{Taylor1,AND,refine}.  In off-grid estimation, the estimated DOAs are not restricted to a specific grid and can take any value in the continuous parameter space.
The main approaches for off-grid recovery proposed  include:
\begin{enumerate}
  \item \emph{Taylor expansion.} In \cite{Taylor1}, the non-linear dependence in the DOA parameter is linearized via a first order Taylor  series expansion, resulting in a formulation from which their values can be recovered without discretization. However, this kind of method heavily depends on the accuracy of the expansion.
  \item \emph{Atomic norm denoising.} In atomic norm denoising methods, the sparse signals are recovered by solving an  atomic norm based objective function \cite{AND}. The function then can be converted into a semi-definite program that  is solved by off-the-shelf solvers in an  off-grid manner. However, solving the atomic norm objective becomes computationally complex for large scale problems, rendering it infeasible for our mmWave massive MIMO channel estimation problem, in which the dimensionality of the  multivariate quantities tends to be very large.
  \item \emph{Grid refinement.} The idea of grid refinement  was first introduced by Malioutov \emph{et al.} \cite{refine} to mitigate the effect of grid mismatch in DOA estimation. This approach adaptively refines the grid around  candidate spatial locations with a predefined resolution.
      This approach suffers from two main drawbacks: First, the computational complexity grows proportionally with the desired accuracy; Second, since the resolution of the points is increased iteratively, this approach tends to converge to local optimal points, degrading its performance.
\end{enumerate}

Here, we propose an algorithm for recovering the DOAs which is
inspired by grid refinement, while avoiding its drawbacks in terms of recovery performance and computational complexity.
Our method starts with a fixed known dense grid, for which \eqref{block_3} represents a linear transformation of an unknown block sparse vector $\bm x$. The algorithm then alternates as follows: first, for a fixed grid, it recovers a block-sparse $\bm x$; 
then, for  fixed block-sparse $\bm x$, we adjust the angle grid and accordingly the projection matrix $\bm D_{bs}$ to further minimize the cost. By repeating these steps iteratively, we are able to recover DOA angles  which minimize the cost function without necessarily lying on the original grid.

Similarly to grid refinement, our approach changes the  grid iteratively. However, while grid refinement  modifies the resolution around a set of observed points, our algorithm adjusts the grid values in a continuous manner and reduces the number of grid points iteratively. The benefits of this approach over conventional grid refinement are numerically demonstrated in our simulation study in Section V.

To explain the algorithm in detail, define $L_{\phi}^{(0)}$ as the initial guess of the number of unknown paths that will be gradually decreased and tuned during the estimation procedure.
With a slight abuse of notation, we use $\bm \phi =[\phi_1,\cdots,\phi_{L_{\phi}^{(0)}}]$ as the off-grid DOAs to be estimated. Since we do not know the number of paths in advance, $L_{\phi}^{(0)}$ is set initially to a relatively large number, and thus estimating $\bm \phi$  can  be formulated as a sparse signal recovery  problem with an unknown parametric dictionary $\bm D_\text{bs}(\bm \phi)$. In this framework, the objective is not only to estimate the sparse signal, but also to optimize/refine the angle grid such that the parametric dictionary  approaches the true sparsifying dictionary.

To proceed, we recall the definition of the block $l_0$-norm:
\begin{defn}
The $T$-block $\ell_o$-norm of a $T L \times 1$ vector ${\bm x} = [{\bm x}^T[1], \ldots, {\bm x}^T[L]]^T$ is defined as \cite{eldar}
\begin{align}\label{p-sparse}
\| \bm x \|_{0,T}\triangleq\sum_{i=1}^{L}\mathcal I(\| \bm x[i] \|_{2}>0),
\end{align}
where $\mathcal I(\| \bm x[i] \|_{2}>0)$ is an indicator function which equals $1$ if $\| \bm x[i] \|_{2}>0$ and $0$ otherwise, and $\bm x[i]$ is the $i$th block of $\bm x$ containing $T$ consecutive elements. Note that the $T$-block $\ell_o$-norm with $T=1$ reduces to the  conventional $\ell_o$-norm.
\end{defn}

\vspace{-1mm}
Using \eqref{block_3}, we  formulate the DOA estimation problem  for a fixed known grid $\bm \phi$ exploiting the prior knowledge of $\bm x$ being block sparse as
\begin{align}\label{lo_norm}
& \min_{\bm{x}} \| \bm x \|_{0,T} \notag\\
            & \text {s. t. }  \| \text{vec}(\bm Y_{\mathcal N}^T) - \bm D_\text{bs}(\bm \phi) {\bm x}\|_2\le \xi,
\end{align}
where $\xi$ is an error tolerance parameter that is related to the noise statistics.
To recover an off-grid estimate of the DOAs, we first recast the problem  (31) as an  iterative reweighted least squares objective, as in, e.g., \cite{fangjun}. Then, we use the resulting objective to tune the grid vector $\bm \phi$, supporting an off-grid estimate in a computationally feasible fashion.

To formulate the iterative algorithm, let $\bm x^{(\omega)}$ and $\bm{\phi}^{(\omega)}$ be the estimations of $\bm x$ and $\bm \phi$ at the $\omega$th iteration, respectively. At each iteration we form the following $\small{ L_{\phi}^{(\omega)}T \times {L_{\phi}^{{(\omega)}}}T}$ matrix:
\vspace{1mm}
\begin{align}\label{GG}
\!\!\!\bm{G}^{(\omega)}\! \triangleq\! \textup{diag}\!\!
\begin{bmatrix}\frac{1}{(\| \bm x^{(\omega)}[1] \|_2^2)+\epsilon}, \!\!\!& \dots \!\!\!\!& ,\frac{1}{(\| \bm x^{(\omega)}[{L_{\phi}^{(\omega)}}] \|_2^2)+\epsilon}\end{bmatrix}\!\!\otimes \!\bm I_T,
\end{align}
where $\epsilon >0$ is a positive parameter ensuring that $\bm{G}^{(\omega)}$ is well-defined, and ${L_{\phi}^{(\omega)}}$ is the number of unknown angles at the $\omega$th iteration.
The block-sparsity problem (31) is then recast as a reweighed least squares objective problem:
\begin{align}\label{N-change}
\min_{\bm{x}} \bm{x}^H\bm{G}^{(\omega)}\bm{x}+\lambda^{(\omega)} \|\text{vec}(\bm Y_{\mathcal N}^T) - \bm D_\text{bs}\big(\bm {\phi}^{(\omega)}\big) {\bm x}\|_2^2.
\end{align}
The objective in \eqref{N-change} consists of two terms: the weighted norm
$\bm x^H \bm {G}^{(\omega)}\bm x$, which controls the level of block sparsity of the recovered vector $\bm x$, and the term $\|\text{vec}( {\bm Y}_{{\mathcal N}}^T) - \bm D_\text{bs}(\bm {\phi}^{(\omega)}) {\bm x}\|_2^2$, which represents the accuracy of the estimation. The balance between the two terms is controlled by the regularization parameter $\lambda^{(\omega)}$  which we set to
\begin{align}\label{lamd}
\lambda^{(\omega)}   = \frac{ MT}{\| \text{vec}({\bm Y}_{{\mathcal N}}^T) - \bm D_\text{bs}\big(\bm {\phi}^{(\omega)}\big) {\bm x^{(\omega)} }\|_2^2}.
\end{align}

For a given $\bm{\phi}^{(\omega)}=\bm{\phi}$, the optimal value of $\bm{x}$ in \eqref{N-change}  is
\begin{align}\label{x_esti}
\bm{x}^{(\omega+1)}|\bm{\phi}=\big(\bm D_\text{bs}^H(\bm \phi)\bm D_\text{bs}(&\bm\phi)+\big(\lambda^{(\omega)}\big)^{-1}\bm{G}^{(\omega)}\big)^{-1}\times
\notag\\
&\bm D_\text{bs}^H(\bm \phi)\text{vec}({\bm Y}_{{\mathcal N}}^T).
\end{align}
Substituting $\bm{x}^{(\omega+1)}|\bm{\phi}$ back into \eqref{N-change},  we can optimize the grid $\bm{\phi}^{(\omega)}$  in light of the objective \eqref{N-change}  by minimizing
\begin{align}\label{f_theta}
v(\bm{\phi}^{(\omega)})\triangleq
&-\text{vec}({\bm Y}_{{\mathcal N}}^T)^H   \bm D_\text{bs}(\bm \phi)   \big(\bm D_\text{bs}^H(\bm \phi^{(\omega)}) \bm D_\text{bs}(\bm \phi^{(\omega)})\notag\\
&+\big(\lambda^{(\omega)}\big)^{-1}\bm{G}^{(\omega)}\big)^{-1}  \bm D_\text{bs}^H(\bm \phi
^{(\omega)})  \text{vec}({\bm Y}_{{\mathcal N}}^T).
\end{align}
Since directly minimizing \eqref{f_theta} is computationally complex, we propose to gradually decrease the surrogate objective by selecting $\bm\phi^{(\omega+1)}$ that satisfies $ v(\bm\phi ^{(\omega+1)}) \le v(\bm\phi ^{(\omega)})$ for the next iteration.
Since $v(\bm \phi^{(\omega)})$ is differentiable with respect to $\bm{\phi}$, the $(\omega\!+\!1)$th  estimation can be obtained by the  gradient descent method:
\begin{align}\label{theta_gad}
\bm \phi^{(\omega+1)} = \bm{\phi}^{(\omega)} -  u \frac{\partial v(\bm{\phi}^{(\omega)})}{\partial \bm{\phi}},
\end{align}
where $u $ is the step size, and the derivative expression is given in closed-form in the Appendix. The update rule in \eqref{theta_gad} implies that, even if $\bm{\phi}^\text{ul}$ is initialized to a large grid, the iterative algorithm allows the updated DOA estimates to deviate from this initial grid, resulting in off-grid estimation.
\begin{algorithm}[tb]
{\small{
\caption{: DOA Estimation }
\begin{itemize}
\item \textbf{Initialize:} Input ${\bm Y}_{{\mathcal N}}$, ${\bm{x}}^{(0)}=\bm 0, {\bm{\phi}}^{(0)}=\bm 0$, $L_{\phi}^{(0)}$, and $\lambda ^{(0)}$. Iteration index $\omega=0$.
\item \textbf{Step 1:} At iteration $\omega$, based on the previous results ${\bm{x}}^{(\omega)}, \lambda ^{(\omega)}$, refresh $\bm{G}^{(\omega)}$ in \eqref{GG} and construct the function
    $v(\bm \phi^{(\omega)})$ via \eqref{f_theta}.
\item \textbf{Step 2:} Update ${\bm{\phi}}^{(\omega+1)}$  via \eqref{theta_gad}.
\item \textbf{Step 3:} Compute  ${\bm{x}}^{(\omega+1)}$ via \eqref{x_esti} and $\lambda ^{(\omega+1)}$ via \eqref{lamd}.
\item \textbf{Step 4:} If $\|\bm { x}[i]\|^2_2 <\mu$, then remove it, and delete the relevant angle $ \phi_i^{(\omega+1)}$ via \eqref{Lk}.
\item \textbf{Step 5:} Update $L_{\!\phi}^{(\omega+1)}$ via \eqref{Lk}.
\item \textbf{Step 6:} If $\|  {\bm x}^{(\omega+1)}- {\bm{x}}^{(\omega)} \| _2 \ge \eta $, then set $\omega\leftarrow \omega+1$ and go to step 1, otherwise stop.
\end{itemize}}}
\end{algorithm}

In the proposed algorithm, the main complexity lies in calculating $\bm{x}^{(\omega+1)}|\bm{\phi}$ and the first derivative $\frac{\partial v(\bm{\phi}^{(\omega)})}{\partial {\bm\phi}}$. To reduce the computational complexity, a pruning method is introduced: for every $i=1,2,\ldots,L_{\phi}^{(\omega)}$, if $\|\bm x^{(\omega+1)}[i] \|_2^2$ is smaller than some fixed threshold $\mu$, then we delete $\bm x^{(\omega+1)}[i]$ from the vector $\bm x^{(\omega+1)}$, and correspondingly delete the angle $ \phi_i^{(\omega+1)}$ from the vector $\bm\phi^{(\omega+1)}$. We then set $L_{\phi}^{(\omega)}$ to be the length of the preserved $ \bm \phi^{(\omega+1)}$:
\begin{align}\label{Lk}
\hspace{-0.2cm}
\left\{
\hspace{-0.1cm}
\begin{array}{lr}
  \bm{x}_*^{(\omega\!+\!1)}\!=\!\bm {x}^{(\omega\!+\!1)}(\| \bm x^{(\omega\!+\!1)}[i] \|_2^2 >\mu),\quad
   i\!=\!1,\ldots\!,\!L_{\phi}^{(\omega)} \\
  \bm{\phi}_*^{(\omega\!+\!1)}\!=\!\bm{\phi}^{(\omega\!+\!1)}(\| \bm x^{(\omega\!+\!1)}[i] \|_2^2 >\mu), \quad i\!=\!1,\ldots\!,\!L_{\phi}^{(\omega)}  \\
 L_{\phi}^{(\omega\!+\!1)}\!=\!\text{Length}\big(\bm\phi_*^{(\omega+1)}\big),
\end{array}
\right.
\end{align}
where $(\cdot)_*$ represents the preserved vector at each iteration.

When the iterative algorithm satisfies its termination criterion, i.e., $\|  {\bm x}^{(\omega+1)}- {\bm{x}}^{(\omega)} \| _2 $ is less than  some predefined $\eta $, the number of paths $P$ can be estimated using the number of non-zero block values of $\bm x $.
The proposed algorithm is summarized as \textbf{Algorithm 1}.

\vspace{-0.2cm}
\subsection{Delay Estimation Algorithm}
\vspace{-0.1cm}
We next consider the recovery of the delays of each path.
%
Note that \eqref{ofdm h-p} implies that the delay factor $ \tau_p^\text{ul} $ has a  Vandermonde vector $\bm b(\tau_p^\text{ul})$  appearing in the expression for each row of $\bm H_{\mathcal N}^\text{ul}$.
Consequently, we can use the same block structure to estimate the delay  by vectorizing ${{\bm Y}_{{\mathcal N}}}$.
Similarly,  we use $\bm{\tau}\triangleq[\tau_1,\tau_2 ,\cdots,\tau_{L_{\tau}^{(0)}} ]$,
as the off-grid delays to be estimated, where  $\tau_i =\frac{i}{L_{\tau}^{(0)}Nf_o}$, $i\in \{ 1,2,\cdots, {L_{\tau}^{(0)}}\}$, and $L_{\tau}^{(0)}$ is the initial number of unknown delays.
Thus, the sensing matrix for delay estimation can be formulated similarly to \eqref{block_1}. To that aim, define
\begin{align}\label{D_tau}
\!\!\!\bm D_t(\bm{ \tau} )\!&\triangleq[\bm b(\tau_1 )\otimes \bm I_M, \bm b(\tau_2 )\otimes \bm I_M, \cdots ,\bm b(\tau_{L_{\tau}}^{(0)} )\otimes \bm I_M ] \notag\\
     \!&=[\bm d_t({\tau_1 }),\bm d_t({\tau_2 }),\cdots,\bm d_t({\tau_{L_{\tau}^{(0)}} })],
\end{align}
where $\bm d_t({\tau_i })\triangleq \bm b(\tau_{i} )\otimes \bm I_M$.
It can be readily checked that
\begin{align}
\text{vec}({\bm Y}_{{\mathcal N}})\approx \bm D_t(\bm{\tau} ) \bm x_t +\vect (\bm E_{\mathcal N}),
\end{align}
where $\bm x_t$ is an $ML_{\tau}^{(0)}\times 1$ block sparse vector  defined similarly to the vector $\bm x$ introduced in the previous subsection.

For a fixed delay grid $\bm \tau$, the delay estimation can be expressed as
\vspace{-1mm}
\begin{align}
&  \min_{\bm x_t} \| \bm x_t \|_{0,M} \notag\\
            &\textup {s. t. }  \| \text{vec}({\bm Y}_{{\mathcal N}}) - \bm D_t(\bm {\tau} ) {\bm x_t}\|_2\le \xi_t,
\end{align}
which can be solved in a similar manner as \eqref{lo_norm}, namely, by iteratively updating the estimate of the vector $\bm x_t $ and the grid $\bm \tau$ using a block-sparsity boosting iterative reweighted least squares objective.
The proposed delay estimation procedure and algorithm is summarized in \textbf{Algorithm  2}.
\begin{algorithm}[tb]
\caption{: Delay Estimation}
{\small{
\begin{itemize}
\item \textbf{Initialize:} Input ${\bm Y}_{{\mathcal N}}$, ${\bm{x}_t}^{(0)}=\bm 0,  \bm{\tau} ^{(0)}=\bm 0$, $L_{\tau}^{(0)}$, and $\lambda ^{(0)}$. Iteration index $\omega_t=0$.
\item \textbf{Step 1:} At iteration $\omega_t$, based on the previous  results $ {\bm{x}_t}^{(\omega_t)}, \lambda ^{(\omega_t)}$, refresh $\bm{G_t}$ in \eqref{GG} and calculate:
    $v_t(\bm{\tau} ^{(\omega)})=-\text{vec}({\bm Y}_{{\mathcal N}})^H   \bm D_{t}(\bm \tau^{(\omega_t)} ) \big(\bm D_{t}^H(\bm \tau^{(\omega_t)} ) \bm D_{t}(\bm \tau^{(\omega_t)})+\big(\lambda_t^{(\omega)}\big)^{-1}\bm{G_t}^{(\omega_t)}\big)^{-1}  \bm D_{t}^H(\bm \tau^{(\omega_t)} )  \text{vec}({\bm Y}_{{\mathcal N}})$.
\item \textbf{Step 2:} Search $v_t(\bm \tau^{(\omega_t)} )$ with gradient descent method for a new estimation of $ {\bm\tau}^{(\omega_t+1)}$ via
 $\bm { \tau} ^{(\omega_t+1)} = {\bm{ \tau} }^{(\omega_t)} - u \frac{\partial v_t(\bm{\tau} ^{(\omega_t)})}{\partial \bm{\tau} }$.
\item \textbf{Step 3:} Compute $ {\bm x_t}^{(\omega_t+1)}=\big(\bm D_{t}^H
(\bm \tau^{(\omega_t)} )\bm D_{t}(\bm \tau^{(\omega_t)} )+\lambda_t^{-1}\bm{G}_t^{(\omega_t)}\big)^{-1}
\bm D_{t}(\bm \tau^{(\omega_t)} )^H\text{vec}({\bm Y}_{{\mathcal N}})$ and $\lambda_t ^{(\omega_t+1)}$ via \eqref{lamd}.
\item \textbf{Step 4:} If $\| \bm x_{t}^{(\omega_t+1)}[i]\| <\mu$, then remove it, and delete the relevant delay $  \tau_i^{(\omega_t+1)} $.
\item \textbf{Step 5:} Update $L_{\tau}^{(\omega_t+1)}$ based on the new length of $( {\bm{\tau}} )^{(\omega_t+1)}$.
\item \textbf{Step 6:} If $\|   {\bm x_t}^{(\omega_t+1)}-  {\bm{x}_t}^{(\omega_t)} \| _2 \ge \eta $, then set $\omega_t\leftarrow \omega_t+1$ and go to step 1, otherwise stop.
\end{itemize}}}
\end{algorithm}

\vspace{-0.2cm}
\subsection{Uplink Channel Reconstruction}
\vspace{-0.1cm}
In the previous subsections we showed how the estimations of the DOAs and the delays, denoted $\bm {  \hat \phi}^\text{ul}$  and $\bm{  \hat\tau}^\text{ul}$  are separately obtained, along with the number of paths $\hat P$.  Yet, one still has to match each DOA value to its corresponding delay, namely, to match the entries of  $\bm { \hat \phi}^\text{ul}$ to their corresponding entries
$\bm{ \hat\tau}^\text{ul}$. A trivial approach is to search all of the $\hat P^{\hat P}$  possible pairings among $\bm {\hat\phi}^\text{ul}$ and $\bm{\hat\tau}^\text{ul}$, but this would incur a heavy computational burden.
Fortunately, based on Theorem 1, if $ \hat\phi_{i}^\text{ul}$  and $\hat\tau_{j}^\text{ul}$ belong to the same path, then the projection of their channel vector $\text{vec}(\bm {\Xi}(\hat\phi_{i}^\text{ul},\hat\tau_{j}^\text{ul}))$ onto $\text{vec}({\bm Y}_{{\mathcal N}})$ will be a larger value compared to other (mismatched) combination.

Consequently, we can try different combinations of $\text{vec}(\bm {\Xi}(\hat\phi_{i}^\text{ul},\hat\tau_{j}^\text{ul}))$ to perform the inner product operation with $\text{vec}({\bm Y}_{{\mathcal N}})$, and select the maximum value from each operation as the correct matching of the angle $ \hat\phi^{\text{ul}}$ for delay $ \hat\tau^\text{ul}$. Next, we delete the angle and delay that have been matched already, and perform the inner product for the remaining angles and delays. Under this pairing procedure, at most $\sum_{p=1}^{\hat P }p^2 =\hat P ^2$ inner product operations are required.

After the pairing process is concluded, one can recover the complex channel gains. To that aim,  we stack the obtained channel vectors (with correct matching) $\vect(\bm {\Xi}(\hat\phi_{p}^\text{ul},\hat\tau_{p}^\text{ul}))$, where $1\le p\le \hat P $, as columns to form a matrix:
\begin{align}
 \bm B\!\triangleq\![\vect(\bm {\Xi}(\hat\phi_{1}^\text{ul},\hat\tau_{1}^\text{ul})),\cdots\!,\vect(\bm {\Xi}(\hat\phi_{\hat P }^\text{ul},\hat\tau_{\hat P }^\text{ul}))].
\end{align}
Now, we note that if there is no error in the estimation of the DOAs and the delays, then the true channel gains can be computed via $\bm \beta = \bm B ^\dagger \vect(\bm H^{\text{ul}})$.  However, as we do not have access to the true channel $\bm H^{\text{ul}}$, we estimate the gains by applying the same transformation to the least-squares estimation of $\bm H^{\text{ul}}$, i.e., $\bm {\bm Y}_{\mathcal N}(\bm S^{\text{ul}})^{-1}$. Specifically, the uplink complex channel gains are estimated as
\begin{align}
 \bm {\hat \beta}^\text{ul}=\bm B ^\dagger \vect\big({\bm Y}_{\mathcal N}(\bm S^{\text{ul}})^{-1}\big)=[{\hat \beta}_{1}^\text{ul},{\hat \beta}_{2}^\text{ul},\ldots,{\hat \beta}_{\hat P }^\text{ul}]^T.
\end{align}
The overall uplink channel at all $N$ subcarriers of the $k$th user can be reconstructed from \eqref{ofdm h-p} as
\begin{align}\label{h-recover}
\bm {\hat H}_k^\text{ul} &= \sum_{p=1}^{\hat P } {\hat \beta}_{p}^\text{ul} \bm {\Xi}(\hat\phi_{p}^\text{ul},\hat\tau_{p}^\text{ul}).
\end{align}
We summarize the overall proposed uplink channel reconstruction scheme as \textbf{Algorithm 3}.
 \begin{algorithm}[tb]
  \caption{Uplink Channel Reconstruction Scheme}
  \label{alg:channel}
  {\small{
  \begin{algorithmic}[1]
    \Require
      The BS receives matrix ${\bm Y}_{{\mathcal N}} $ via \eqref{Y}.
    \Ensure
      The estimated channel matrix $ \bm {\hat H}_k^\text{ul}$ over $N$ subcarriers,
      Number of paths $\hat P$, DOAs $\bm {\hat\phi}^\text{ul}$, delays $\bm {\hat\tau}^\text{ul}$, and complex gain $\bm {\hat \beta}^\text{ul}$ for each user.

     \State Calculate the DOAs $ \phi_{i}, 1\le i \le \hat P , $ using \textbf{Algorithm 1} with input ${\bm Y}_{{\mathcal N}}$.

    \State Calculate the delays $ \tau_{j}, 1\le j \le \hat P $, using  \textbf{Algorithm 2} with input ${\bm Y}_{{\mathcal N}}$.

    \While{$p<\hat P $ }
                    \State $\text {value} \gets \text{zeros}[\hat P -p+1,\hat P -p+1]$
                    \For{$j = p : \hat P $}
                    \State $\text {value}[p,j] \gets \vect(\bm {\bm Y}_{{\mathcal N}})^H \vect(\bm {\Xi}(\phi_{p},\tau_{j})) $
                    \EndFor
                    \State $[pos_{p},pos_{j}] \gets \bf{find}(\bf {max}(\text{value})) $
                    \State $\hat\phi_{p}^\text{ul}=\phi_{pos_{p}}; \hat\tau^\text{ul}_{p}=\tau_{pos_{j}}$
                    \State $p \gets p+1$

    \EndWhile
    \State Stack  $\vect(\bm {\Xi}(\hat\phi_{p}^\text{ul},\hat\tau_{p}^\text{ul}))$ into the columns of the matrix:
     $\bm B_k$=$[\vect(\bm {\Xi}(\hat\phi_{1}^\text{ul},\hat\tau_{1 }^\text{ul})),\cdots ,\vect(\bm {\Xi}(\hat\phi_{\hat P }^\text{ul},\hat\tau_{\hat P }^\text{ul}))]$.
    \State Calculate the complex gain as $\bm {\hat \beta}=\bm B^\dagger \vect({\bm Y}_{\mathcal N}(\bm S^{ul})^{-1})$.
    \State Recover the channel matrix  $\bm{\hat H}_{k}^\text{ul}$ via \eqref{h-recover}. \\
    \Return $\bm {\hat\phi}^\text{ul},\bm {\hat\tau}^\text{ul},\bm {\hat\beta}^\text{ul},\bm {\hat H}_k^\text{ul}$.
  \end{algorithmic}}}
\end{algorithm}


\vspace{-0.2cm}
\section{ Downlink  Channel Estimation }
\vspace{-0.1cm}
In the previous section we proposed an algorithm for estimating the uplink channel.  However, in order to establish reliable bi-directional communications, the downlink channel must also be estimated. One of the major challenges in massive MIMO communications stems from the fact that in FDD systems, in which different bands are assigned to uplink and downlink transmissions, the downlink channel cannot be immediately deduced from the uplink channel. In the following we show how, for mmWave massive MIMO systems with the BSE, the estimation scheme designed for uplink channels in the previous section can be extended to downlink channels by exploiting a phenomenon called {\em angle-delay reciprocity}.

In particular, we consider pilot-aided downlink channel estimation in which the BS transmits a-priori known pilot sequence in a similar manner to the uplink channel estimation phase.
To deal with the BSE, we design the downlink estimation strategy to use dedicated pilots for each path and using different beamforming vectors for different subcarriers with respect to the given path.
To present our scheme, we first elaborate on the structure of mmWave massive MIMO downlink channels in the presence of BSE in Subsection IV-A. Then, in Subsection IV-B, we discuss the angle-delay reciprocity, a property which we exploit in Subsection IV-C to generate the beamformed pilots  and to formulate our downlink channel estimation algorithm.

\vspace{-0.2cm}
\subsection{Downlink Channel Structure}
\vspace{-0.1cm}
To formulate the downlink channel, we denote its  center frequency $f_c^\text{dl}$ and wavelength $\lambda^\text{dl}={c}/{f_c^\text{dl}}$. The downlink array steering vector is expressed as
\begin{align}\label{a_theta_dl}
\!\!\bm a^\text{dl}(\phi^\text{dl},f)\!\!=\!\Big[1,e^{-j (1+\frac{f}{f_c^\text{dl}})\phi^\text{dl}},\cdots,e^{-j (M-1)(1+\frac{f}{f_c^\text{dl}})\phi^\text{dl} }\Big]^T\!\!\!,
\end{align}
where $\phi^\text{dl} \triangleq \frac{2\pi d\cdot \sin \theta^\text{dl}}{\lambda^\text{dl}}$ is  the normalized direction of departure (DOD), and  $\theta^\text{dl}$ is the downlink DOD. We henceforth use the term ``DOD" to refer to the normalized DOD $\bm \phi^{\text{dl}}$.

Similar to the uplink case, at frequency $f$, the downlink channel observed by the $k$th user can be written as:
\begin{align}\label{channel hf_down}
 \bm h_{k}^\text{dl}(f)= \sum_{p=1}^{ P^{\text{dl}} } \bar \beta_{p}^\text{dl} \bm a^\text{dl}(\phi_{p}^\text{dl},f) e^{-j2\pi f\tau_{p}^\text{dl}},
\end{align}
where $P^{\text{dl}}$ is the number of downlink paths, $\bar \beta_{p}^\text{dl}$  and $\tau_{p}^\text{dl}$ are the complex gain and  multi-path delay of the $p$th downlink path, respectively. The channel parameters, $P^{\text{dl}}$ , $\bar \beta_{p}^\text{dl}$, $\tau_{p}^\text{dl}$, and $\phi_{p}^\text{dl}$  depend on the specific user index $k$, which is omitted  for brevity,  as in the previous section.

As in \eqref{W_mn} and \eqref{ofdm h-p}, we define :
\begin{align}
\big[\bm W^\text{dl}(\phi^\text{dl}_{p})\big]_{m,n} &\triangleq\text{exp}\big(-j m\frac{nf_0}{f_c^\text{dl}}\phi_{p}^\text{dl}\big); \notag\\
 \bm {\Xi}^\text{dl}(\phi^\text{dl}_{p},\tau^\text{dl}_{p})&\triangleq\left[\bm a(\phi^\text{dl}_{p},0) \bm b^T(\tau^\text{dl}_{p})\right] \odot \bm W^\text{dl}(\phi^\text{dl}_{p}).
\end{align}
Then, the downlink $1\times MN$ frequency channel vector from the BS to the $k$th user is given by
\begin{align}\label{channel H_down}
\bm H_{k}^\text{dl} &=\left[\vect \left[\bm h_{k}^\text{dl}(0),\bm h_{k}^\text{dl}(f_0), \cdots ,\bm h_{k}^\text{dl}((N-1)f_0) \right]\right]^H\notag\\
                   &=\!\!\sum_{p=1}^{ P^{\text{dl}} }\!\! \Big[{\bar \beta}_{p}^\text{dl} \vect \left( \left[\bm a^\text{dl}(\phi_{p}^\text{dl},0) \bm b^T(\tau_{p}^\text{dl})\right] \!\!\odot \!\!\bm W^\text{dl}(\phi_{p}^\text{dl})  \right )\Big]^H \notag\\
                  &=(\bm { \beta}^\text{dl})^H \bm \Omega_k(\bm \phi^\text{dl},\bm \tau^\text{dl})^H  ,
\end{align}
where $\bm\beta^\text{dl}=[\bar \beta_{1}^\text{dl},\bar \beta_{2}^\text{dl},\ldots,\bar \beta_{ P^{\text{dl}}}^\text{dl}]^T $ is the downlink channel complex gain vector, and
\begin{align}\label{downlink}
\!\!\!\!\!\!\bm \Omega_k(\bm \phi^\text{dl},\!\bm \tau^\text{dl}\!)\!\!= \!\![\vect\!\left(\bm {\Xi}^\text{dl}(\phi_{1}^\text{dl},\tau_{1}^\text{dl})\!\right),\!\ldots\!,\vect\!\left(\bm {\Xi}^\text{dl}(\phi_{P^{\text{dl}} }^\text{dl},\tau_{{P^{\text{dl}} }}^\text{dl})\!\right)\! ].
\end{align}

From \eqref{channel H_down}, the downlink channel consists of a set of DODs, delays, and complex gains, and thus has a similar structure as the uplink case.
Note that in FDD systems,   channel reciprocity does not hold, and the channel must be estimated independently by each user. However, this estimation can be facilitated by accounting for the angle-delay reciprocity of mmWave channels, presented in the following subsection.

\vspace{-2mm}
\subsection{Angle-Delay Reciprocity}
\vspace{-0.1cm}
Unlike in TDD systems \cite{channel-recip}, FDD channels are not reciprocal, namely, downlink and uplink transmissions undergo different channels due to their different frequency bands.
However, since the propagation paths of electromagnetic waves are reciprocal, only the signal wave that physically reverses the uplink path can reach users during  downlink transmission. It has been shown in \cite{angle-delay1,angle-delay2,huyi1} that the conductivity and relative permittivity of most materials
remain unchanged if the frequency of the electromagnetic wave does not vary much, say less than 1GHz.
Hence, the angle components of the uplink  and downlink channels commonly are the same  in mmWave communications \cite{hongxiang}.
Moreover, since the downlink electromagnetic wave travels the same distance as the uplink, the delay components for the uplink and downlink channel are the same.
This phenomenon is commonly referred to as  \textit{angle-delay reciprocity} \cite{bolei-1}, and it implies that
\begin{align}\label{recip}
  \theta_{p}^\text{dl} =  \theta_{p}^\text{ul},\quad
  \tau_{p}^\text{dl} =  \tau_{p}^\text{ul},\quad
  P^{\text{dl}}=   P.
\end{align}
From \eqref{recip}, the uplink and downlink channels have the same number of paths as well as the same angle and delay parameters, which can be estimated at the BS.
Therefore, to acquire the downlink channel in FDD systems, the users only need to estimate the remaining downlink channel gain and feedback this gain to the BS. The resulting  computational overhead, as we show in the sequel, can be made affordable.

\vspace{-0.2cm}
\subsection{Downlink Channel Estimation }
\vspace{-0.1cm}
We now show how the angle-delay reciprocity can be exploited to estimate the downlink channel. Here, the BS estimates \eqref{recip}, using the number of uplink paths along with their DOAs and delays obtained via \textbf{Algorithm} 3.

Recall that the BS transmits beamformed pilots to each user during downlink channel estimation. In particular,
the BS sends a-priori known pilots in each estimated path. Let  $\bm s^\text{dl}_{p} \in \mathbb C^{1\times T}$  denote the pilots targeting the $p$th path over $T$ subcarriers with index set  $\mathcal{N} = \{ n_1, n_2, \ldots, n_T \}$.
These pilots are orthogonal over the different paths, i.e., $\bm s^\text{dl}_{i} (\bm s^\text{dl}_{j})^H=\delta(i-j)$.
To formulate how these pilots are beamformed prior to their transmission, we use the channel structure in \eqref{channel H_down} and formulate the columns of the matrix $\bm \Xi^{^{\mathcal{N}}}(\phi_{p}^\text{dl},\tau_{p}^\text{dl})$ as follows: for the $n_q$th ($q=1,2,\cdots,T$) pilot subcarrier, the corresponding column of $\bm \Xi^{\mathcal{N}}(\phi_{p}^\text{dl},\tau_{p}^\text{dl})$ is given by
\begin{align}
\!\!\bm {\Xi}_{:,q}^{\mathcal{N}}(\phi_{p}^\text{dl},\tau_{p}^\text{dl})\!\!=\!\! \left[\bm a(\phi_{p}^\text{dl},0) \bm b_{n_q}^T(\tau_{p}^\text{dl})\right]\!\!\odot\![\bm W^{\text{dl}}(\phi_{p}^\text{dl})]_{:,{n_q}},
\end{align}
where {\small{$\bm b_{n_q}^T(\tau_{p}^\text{ul})$}} is the $n_q$th element of
{\small{$\bm b^T(\tau_{p}^\text{dl})$}}, and {\small{$ [\bm W^{\text{dl}}(\phi_{p}^\text{ul})]_{:,{n_q}}$}} is the $n_q$th column of {\small{$\bm W^{\text{dl}}(\phi_{p}^\text{ul})$}}.
By letting {\small{$\bm F_{{n_q}}^p\big(\phi_{p}^\text{dl},\tau_{p}^\text{dl}\big)$}} be the  beamforming vector  for the $k$th user from the $p$th path on the $n_q$th pilot carrier, the corresponding channel output (prior to the addition of noise) can be written as
\begin{align}
r_{{n_q},p}^\text{dl}\!\!=\!\big[{\bar \beta}_{p}^\text{dl}
\vect(\bm {\Xi}_{:,q}^{\mathcal{N}}(\phi_{p}^\text{dl},\tau_{p}^\text{dl}))\big]^H&\bm F_{{n_q}}^p\big(\phi_{p}^\text{dl},\tau_{p}^\text{dl}\big)\bm s_{p}^\text{dl}(q).
\end{align}
In order to point the downlink beam to the $p$th path, we set the beamforming vector to be
\begin{align}
\bm F_{{n_q}}^p\big(\phi_{p}^\text{dl},\tau_{p}^\text{dl}\big)=\frac{1}{M}\vect\big(\bm {\Xi}_{:,q}^{\mathcal{N}}(\phi_{p}^\text{dl},\tau_{p}^\text{dl})\big).
\end{align}

The ${1\times T}$ received vector from all $T$ pilot carriers can now be written as
\begin{align}
\bm y_{{\mathcal N}}^\text{dl}&=
\begin{bmatrix}
\sum\limits_{p=1}^{ P } r_{{ n_1},p}^\text{dl},&\sum\limits_{p=1}^{ P } r_{{ n_2},p}^\text{dl},&\cdots, &\sum\limits_{p=1}^{ P } r_{{ n_T},p}^\text{dl}
\end{bmatrix}+\bm e^\text{dl}\notag \\
&=\bm H_{{\mathcal N}}^\text{dl} \bm F_{\mathcal N}(\bm \phi^\text{dl},\bm \tau^\text{dl}) \bm S^\text{dl} + \bm e^\text{dl},
\end{align}
where {\small{$\bm H_{{\mathcal N}}^\text{dl}\triangleq \sum_{p=1}^{ P } \big[{\bar \beta}_{p}^\text{dl} \vect(\bm {\Xi}^{\mathcal N}(\phi_{p}^\text{dl},\tau_{p}^\text{dl}))\big]^H $}} is the {\small{${1\times MT}$}} downlink channel over the {\small{$T$}} pilot carriers, {\small{ $\bm S^\text{dl} = [(\bm s^\text{dl}_{1})^T,(\bm s^\text{dl}_{2})^T,\cdots, (\bm s^\text{dl}_{\hat P })^T]^T$}} is the pilot matrix, and {\small{$\bm F_{\mathcal N}(\bm \phi^\text{dl},\bm \tau^\text{dl})$}} is the beamforming matrix given by
\begin{align}\label{BF}
\bm F_{\mathcal N} \!(\bm \phi^\text{dl},\bm \tau^\text{dl}\!)=
\!\!\frac{1}{MT} [\vect(&\bm {\Xi}^{\mathcal N}\!(\phi_{1}^\text{dl},\tau_{1}^\text{dl})),\!\cdots\!,\vect(\bm {\Xi}^{\mathcal N}(\phi_{{\hat P }}^\text{dl},\tau_{{\hat P }}^\text{dl})) ].
\end{align}
Due to the beamforming matrix in \eqref{BF}, the user can use the a-priori knowledge of $\bm S^\text{dl}$ to recover its downlink channel complex gain vector using simple least-squares estimation:
\begin{align}\label{DL-gain}
 (\bm{\hat \beta}^\text{dl})^H=\bm y_{{\mathcal N}}^\text{dl} (\bm S^\text{dl})^{\dag}.
\end{align}
To allow the BS to recover the complete downlink channel, each user  now feedbacks its estimated gain vector $ \bm{\hat \beta}^\text{dl}$ to the BS,  completing the downlink channel reconstruction via:
\begin{align}\label{channel H_down_res}
\bm {\hat H}_{k}^\text{dl} =(\bm {\hat\beta}^\text{dl})^H  \bm \Omega_k(\bm \phi^\text{dl},\bm \tau^\text{dl})^H.
\end{align}

We summarize the  downlink channel reconstruction  scheme as \textbf{Algorithm 4}.
\begin{algorithm}[tb]
  \caption{Downlink Channel Reconstruction Scheme}
  \label{alg:channel}
  {\small{
  \begin{algorithmic}[1]
    \Require
      The uplink channel estimation parameters $\bm {\hat\phi}^\text{ul},\bm {\hat\tau}^\text{ul}, \hat P $ for the $k$th user;
      The downlink orthogonal pilots $\bm S^{\text{dl}}$ for the $k$th user.
    \Ensure
    The reconstruction downlink channel matrix $\bm {\hat H}_{k}^\text{dl}$.
    \State Uplink channel estimation:  the $k$th user sends orthogonal pilots $\bm S^{\text{ul}}$ to BS. The BS uses \textbf{Algorithm 3} for estimating uplink channel, recovering the DOAs, delays, and multi-path numbers.
    \State Downlink training: The BS  generates the beamforming matrix $\bm F_{\mathcal N} \!(\bm \phi^\text{dl},\bm \tau^\text{dl})$ in \eqref{BF}, and sends the pilots $\bm S^\text{dl}$ to the $k$th user. After receiving the pilots, the $k$th user recovers the complex gain $\bm {\hat\beta}^\text{dl}$ via \eqref{DL-gain}.
    \State Downlink channel reconstruction: the $k$th user feedbacks the downlink complex gain to the BS. Then the downlink channel is reconstructed via \eqref{channel H_down_res}\\
    \Return  $\bm {\hat H}_{k}^\text{dl}$.
  \end{algorithmic}}}
\end{algorithm}

\vspace{-0.2cm}
\section{Simulations}
\vspace{-0.1cm}
In this section, we demonstrate the effectiveness of the proposed algorithms for uplink and downlink channel estimation compared to conventional channel estimation algorithms. In particular, we show that our  approach significantly outperforms previously proposed methods, which either restrict the solution set to a finite grid, as in \cite{OnCS1,OnCS2}, or, alternatively, do not take into account the BSE. We also compare our off-grid recovery schemes to conventional grid refinement \cite{refine}.

We consider a BS  equipped with a ULA with element spacing $d=\lambda^\text{ul}/2$.
All $K=8$ users are randomly distributed in the service area and each has a single antenna.
The pilots are uniformly distributed over all the $N=64$ subcarriers, thus each user is assigned $T=8$ pilot subcarriers.
The transmit bandwidth is $1$GHz with uplink center frequency $f_c=60$GHz and downlink center frequency $f_c=61$GHz.
For the on-grid approach, we take $L=1024$ grid points. For the proposed off-grid approach, we correspondingly set $L_{\phi}=1024$ as the initial resolution.
The simulated mmWave channels are generated via \eqref{ofdm h-wide} with $ P =6$, $\bar\beta_p\sim\mathcal{CN}(0,1)$, $\phi_p\sim\mathcal{U}({-\pi},{\pi})$, and $\tau_p\sim\mathcal{U}(0,\frac{1}{Nf_0})$, where $\mathcal{CN}$ and $\mathcal U$ represent the complex normal and the uniform distribution, respectively.
The signal-to-noise ratio (SNR) is defined as $\sigma_p^2/\sigma_n^2$, where $\sigma_p^2$ is the pilot power.
The performance of angle and delay estimation are measured by the corresponding mean-square error (MSE) values, defined as
\begin{align}
\text{MSE}_{\bm \phi}&\triangleq \frac{1}{K J}\sum_{k=1}^{K}\sum_{j=1}^{J}{\| \bm \phi_{k,j}-\bm{\hat \phi}_{k,j}\|_2^2},\\
\text{MSE}_{\bm \tau}&\triangleq \frac{1}{KJ}\sum_{k=1}^{K}\sum_{j=1}^{J}{\| \bm \tau_{k,j}-\bm{\hat \tau}_{k,j}\|_2^2},
\end{align}
respectively. Here, $J = 1000$ is the number of Monte-Carlo trials.
The channel estimation performance is measured in terms of the normalized mean square
error (NMSE):
\begin{align}
\text{NMSE}\triangleq \frac{1}{KJ}\sum_{k=1}^{K}\sum_{j=1}^{J}\frac{\| \bm H_{k,j} -\bm{\hat H}_{k,j}\|_F^2}{\| \bm H_{k,j} \|_F^2}.
\end{align}
\begin{figure}[t]
\centering
\vspace{-0.6cm}
\includegraphics[width=90mm]{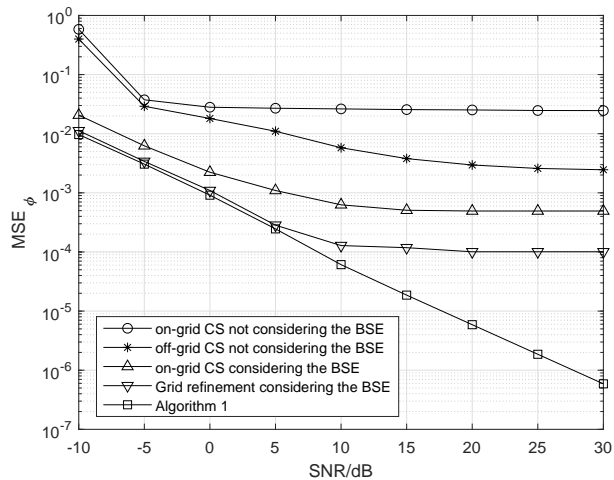}
\caption{MSE vs. SNR, DOA estimation, uplink channel.
\label{fig:theta1}}
\end{figure}

Fig. \ref{fig:theta1} shows the uplink $\text{MSE}_{\bm\phi}$ versus SNR of the proposed algorithm compared with  on-grid CS approach \cite {OnCS2}, grid refinement approach \cite{refine}  and  conventional channel modeling \cite{fangjun} that ignores the BSE.
While the MVV structure is not considered in these algorithms \cite{OnCS2,refine,fangjun}, in the following we allow the competing algorithms to exploit this structure in order to maintain a fair comparison,
Observing Fig. \ref{fig:theta1}, we note that the curve for the MSE in recovering the DOAs of the proposed algorithm decreases linearly as SNR increases (indicating that  $\text{MSE}_{\bm \phi}$ decays exponentially with increasing SNR), while all the other methods meet error floors at high SNR.
This error floor exhibited by previously proposed estimators,  which emphasizes the benefits of our proposed algorithms in high SNR values, is a result of a model mismatch which can be attributed to:
1) For the on-grid algorithms, the grid mismatch restricts the resolution of DOA estimation to be $2\pi/L$;
2) For the grid refinement algorithm,  the error floor  occurs due to   convergence to  local optimum points for some of the parameters, as mentioned in Section III.B.
3) For the off-grid algorithm, neglecting the BSE  significantly decreases  channel estimation accuracy.
It is also observed from the second and the third curves in Fig. \ref{fig:theta1} that  the degradation due to ignoring the BSE is more substantial compared to grid mismatch.

\begin{figure}[t]
\centering
\vspace{-0.6cm}
\includegraphics[width=90mm]{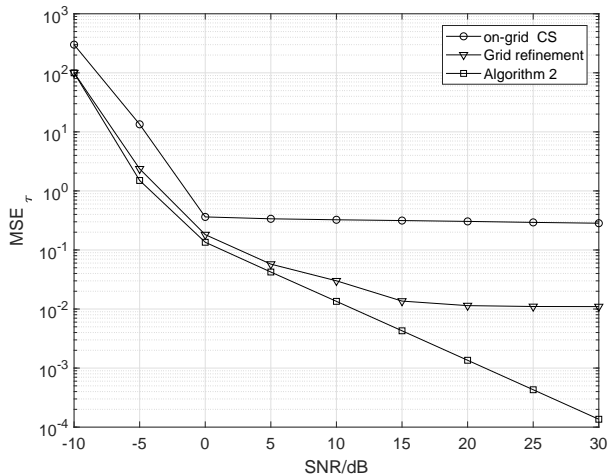}
\caption{MSE vs. SNR, delay estimation, uplink channel.
\label{fig:64-delay}}
\end{figure}

Fig. \ref{fig:64-delay} depicts the MSE in recovering the delays for the same setup.
Since the BSE does not influence the delay estimation, we only present the results for the on-grid, grid refinement, and the proposed off-grid approaches.
Similarly to Fig.~\ref{fig:theta1}, the on-grid  and  grid refinement methods inevitably encounter an error floor at high SNRs, while the proposed off-grid method  consistently improves  performance.
\begin{figure}[t]
\centering
\vspace{-0.6cm}
\includegraphics[width=90mm]{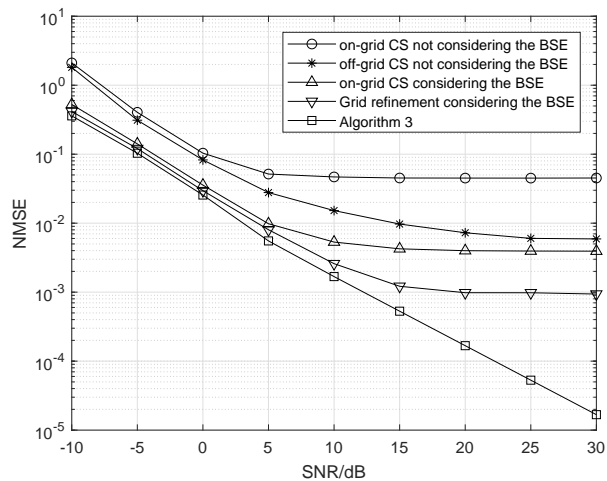}
\caption{NMSE vs. SNR, reconstructed uplink channel.
\label{fig:64-H}}
\end{figure}

Next, we compare the NMSE in recovering the uplink channel using our proposed $\textbf {Algorithm 3}$.
The results are depicted in Fig. \ref{fig:64-H}.
It is observed that the NMSE curve for the proposed algorithm decreases linearly as the SNR increases, achieving significantly better performance than competing methods.
Furthermore, the channel estimation accuracy of the other four techniques all meet error floors at high SNR, in correspondence with their $\text{MSE}_{\bm \phi}$ and $\text{MSE}_{\bm \tau}$ performance.
\begin{figure}[t]
\centering
\vspace{-0.6cm}
\includegraphics[width=90mm]{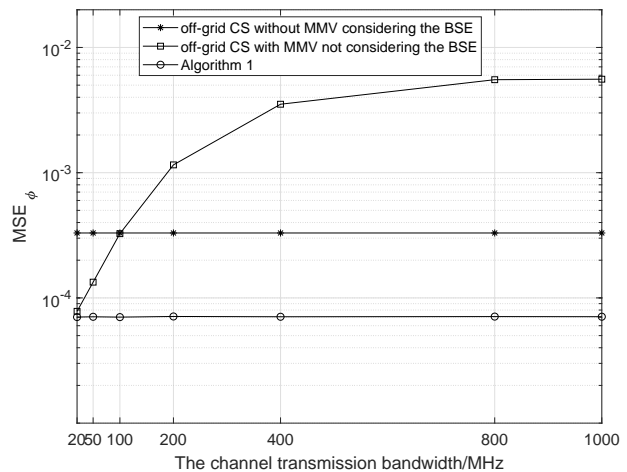}
\caption{MSE vs. bandwidth, DOA estimation, uplink channel.
\label{fig:band-theta}}
\vspace{-0.3cm}
\end{figure}

We next study the effect of  bandwidth on the performance of our estimators. The SNR is set  to 10dB. Fig. \ref{fig:band-theta} depicts the MSE in estimating the DOAs for uplink mmWave massive MIMO channels of the proposed algorithm compared to  off-grid approaches \cite{fangjun} that either ignore the BSE or do not utilize the MMV structure
under various transmission bandwidths.
It is observed in Fig. \ref{fig:band-theta}  that the proposed algorithm achieves the best estimation accuracy and that its superiority over previously proposed estimators is consistent for various bandwidths.
When the bandwidth is as small as $20$MHz, it is noted that the off-grid approach considering the MMV structure but ignoring the BSE  has the same performance as our proposed method. This is because the BSE is not pronounced when the bandwidth is small.
However, as the bandwidth increases, the performance of the algorithms which ignore the BSE  quickly deteriorates.  
Furthermore, the algorithm that utilizes the MMV structure but does not consider the BSE performs worse than the proposed one, and  the performance gap remains constant for different bandwidth values. This demonstrates
that properly exploiting the block sparsity can improve the performance and  this improvement is not affected by bandwidth.

\begin{figure}
\centering
\vspace{-0.6cm}
\includegraphics[width=90mm]{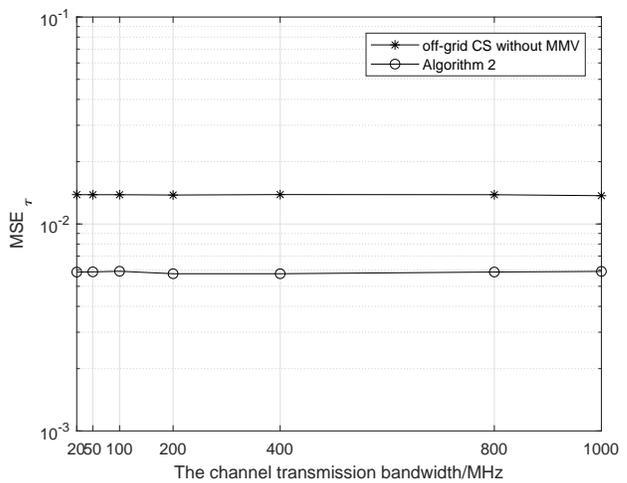}
\caption{ MSE vs. bandwidth, delay estimation, uplink channel.
\label{fig:band-delay}}
\vspace{-0.3cm}
\end{figure}

Fig. \ref{fig:band-delay} displays the uplink MSE in recovering the delays with the same system setup in Fig. \ref{fig:band-theta}.
Because the BSE does not affect the delay estimation, the MSE in estimating the path delays for both algorithms remains constant as the bandwidth increases.
Similarly to Fig. \ref{fig:band-theta}, the proposed algorithm still achieves superior performance for all bandwidth values.
\begin{figure}[t]
\centering
\vspace{-0.5cm}
\includegraphics[width=90mm]{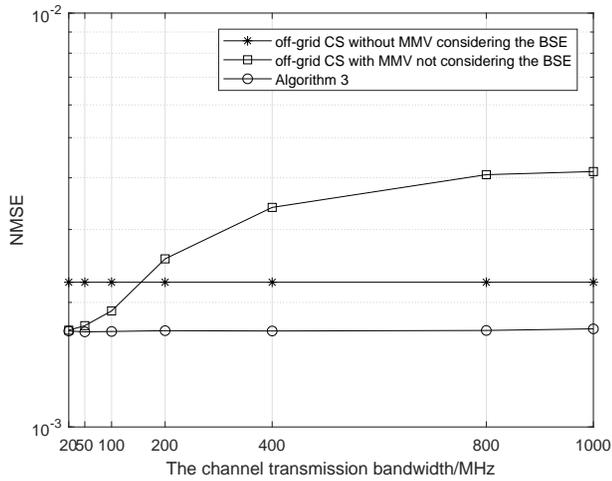}
\caption{NMSE vs. bandwidth,  reconstructed uplink channel.
\label{fig:band-H}}
\end{figure}

Fig. \ref{fig:band-H} compares the NMSE of the reconstructed uplink channel via $\textbf {Algorithm 3}$ with  the competing approaches under various  bandwidths. As the bandwidth increases, we observe that the NMSE of the proposed algorithm remains constant and yields the best performance over all the techniques,  settling with the results depicted in Figs. \ref{fig:band-theta}-\ref{fig:band-delay}.
\begin{figure}[t]
\centering
\vspace{-0.5cm}
\includegraphics[width=90mm]{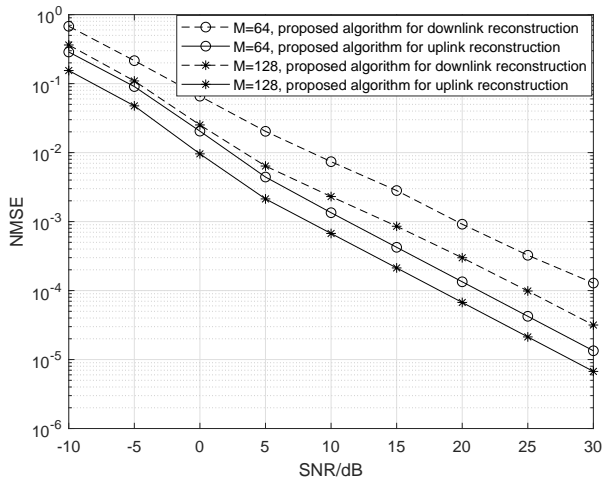}
\caption{NMSE vs. SNR of the reconstructed channel for both uplink and downlink channel under different antenna size.
\label{fig:64-128}}
\end{figure}

In the final example, we  evaluate the NMSE of the reconstructed channel for both the uplink and downlink with different number of antennas  in Fig. \ref{fig:64-128}.
Observing Fig. \ref{fig:64-128}, we note  that the proposed algorithm is very effective in estimating the downlink channel and achieves a consistent improvement in  performance as SNR increases.
Nevertheless, the NMSE in recovering the downlink channel is always worse than that of the uplink channel because the angle and delay parameters of the downlink channel originated from their uplink counterparts and may include estimation errors.
These errors are combined with the estimation errors which arise due to the presence of noisy observations.
Yet, as the number of antennas $M$ increases, the channel recovery performance of both the uplink and downlink improves accordingly, indicating the potential benefits of our channel estimators for massive MIMO systems.

\vspace{-0.2cm}
\section {Conclusions}
\vspace{-0.1cm}
In this paper, we designed channel estimation algorithms under a non-negligible beam squint effect in  mmWave massive MIMO systems.
We first showed that the recovery of the DOAs and delays of each channel path can be represented as an MMV problem using a shift invariant transformation, and developed an algorithm for recovering an off-grid estimate of these parameters. We then showed how these recovered values can be used to estimate the overall downlink channel.
By exploiting the angle-delay reciprocity of mmWave channels, we extended the results   derived for uplink channel estimation to a computationally efficient approach with low  overhead for downlink channel estimation in FDD systems.
Compared to previously proposed channel estimators, which either adopted an on grid approach or, alternatively, did not account for the beam squint modeling, the proposed algorithms  provide significantly better performance.
Numerical simulation results demonstrated the effectiveness of the proposed techniques, and have shown that  properly taking the BSE into account is critical for mmWave massive MIMO systems.

\vspace{-0.2cm}
\begin{appendix}
\vspace{-0.1cm}
In this appendix, we  show how the derivative of $v(\bm \phi)$  in \eqref{theta_gad} can be computed.
The derivative of $v_t(\bm \tau)$ in \textbf{Algorithm 2} is obtained in a similar fashion  and is thus omitted for brevity.
From \eqref{f_theta},
\begin{align}
v(\bm{\phi})=\!-\text{vec}&({\bm Y}_{{\mathcal N}}^T)^H   \bm D_\text{bs}(\bm \phi)   \big(\bm D_\text{bs}^H(\bm \phi) \bm D_\text{bs}(\bm \phi)+\notag\\
&\big(\lambda^{(\omega)}\big)^{-1}\bm{G}^{(\omega)}\big)^{-1}  \bm D_\text{bs}^H(\bm\phi)  \text{vec}({\bm Y}_{{\mathcal N}}^T).
\end{align}
Define:
\begin{align}
\!\!\bm X \!\!\triangleq \!\! \bm D_\text{bs}(&\bm \phi)    \big(\bm D_\text{bs}^H(\bm \phi) \bm D_\text{bs}(\bm \phi)
\!+\!\big(\lambda^{(\omega)}\big)^{\!-1}\!\bm{G}^{(\omega)}\big)^{-1} \! \!\bm D_\text{bs}^H(\bm\phi).
\end{align}
Based on the chain rule, the first derivative of $v(\bm{\phi})$ with respect to $\phi_i$ can be computed as
\begin{align}
\!\!\frac{\partial v(\bm{\phi})}{\partial \phi_i }\!\!=\text{tr}
\Bigg(  \Big(\frac{\partial v(\bm\phi)}{\partial \bm X }\Big)^T  \frac{\partial \bm X}{\partial \phi_i} \Bigg)
\!\!\!+\text{tr}\Bigg(  \Big(\frac{\partial v(\bm\phi)}{\partial \bm X^{\ast} }\Big)^T  \frac{\partial \bm X^\ast}{\partial \phi_i} \Bigg),
\end{align}
where
\begin{align}
\frac{\partial \bm X}{\partial  \phi_i }&=\frac{\partial}{\partial   \phi_i}\Big(\bm D_{bs}(\bm \phi) \big( \bm D_{bs}^H(\bm \phi) \bm D_{bs}(\bm \phi)+\rho \bm I\big)^{-1}\bm D_{bs}^H(\bm \phi)\Big)\notag\\
&=\frac{\partial \bm D_{bs}(\bm \phi)}{\partial  \phi_i}\Big( \big(\bm D_{bs}^H(\bm \phi) \bm D_{bs}(\bm \phi)+\rho \bm I\big)^{-1}\bm D_{bs}^H(\bm\phi)\Big)+\notag\\
&\!\!\!\!\!\!\!\!\Big(\bm D_{bs}(\bm \phi) \big(\bm D_{bs}^H(\bm \phi) \bm D_{bs}(\bm \phi)+\rho \bm I\big)^{-1}\Big)\frac{\partial \bm D_{bs}^H(\bm \phi)}{\partial  \phi_i}-\bm D_{bs}(\bm \phi)\times\notag\\
&\!\!\!\!\!\!\!\!  \big( \bm D_{bs}^H(\bm \phi) \bm D_{bs}(\bm \phi)+\rho \bm I\big)^{-1} \Big( \frac{\partial \bm D_{bs}^H(\phi)}{\partial  \phi_i}\bm D_{bs}( \bm \phi)+ \bm D_{bs}^H( \bm \phi)\times \notag\\
&\!\!\!\!\!\!\!\! \frac{\partial \bm D_{bs}(\bm \phi)}{\partial  \phi_i}\Big)\big( \bm D_{bs}^H(\bm \phi) \bm D_{bs}(\bm \phi)+\rho \bm I\big)^{-1} \bm D_{bs}^H(\bm \phi),
\end{align}
and
\begin{align}
&\frac{\partial v(\bm\phi)}{\partial \bm X  }\!\!=\!\!\frac{\partial}{\partial \bm X  }\text{tr}\!\left(\text{vec}({\bm Y}_{{\mathcal N}}^T\!) \text{vec}({\bm Y}_{{\mathcal N}}^T\!)^H\!\!\bm X\!\right)\!\!=\!\text{vec}({\bm Y}_{{\mathcal N}}^H\!) \text{vec}({\bm Y}_{{\mathcal N}}^H\!),\notag\\
&\frac{\partial v(\bm\phi)}{\partial \bm X^{\ast}  }\!\!=\!\!\frac{\partial}{\partial \bm X ^{\ast} }\text{tr}\left(\text{vec}({\bm Y}_{{\mathcal N}}^T) \text{vec}({\bm Y}_{{\mathcal N}}^T)^H \bm X\right )=0,
\end{align}
where
\begin{align}
\!\!\!\!\!\!\!\!\!\!\!\!\!\!\!\!\!\!  \frac{\partial \bm D_{bs}(\bm \phi)}{\partial  \phi_i}  =\big[\bm 0_{MT,L_{\phi}},\cdots,\frac{\partial \bm D( \phi_i)}{\partial  \phi_i} ,\cdots,\bm 0_{MT,L_{\phi}} \big],
\end{align}
$\bm 0_{MT,L_{\phi}}$ is a zero matrix with dimension of $MT\times L_{\phi}$, and $\bm D( \phi_i)$ is defined in \eqref{Dbi}.

\end{appendix}

\end{document}